\documentclass[aps,pra,twocolumn,twoside,a4paper,10pt,amsmath,amssymb,superscriptaddress]{revtex4-1}

\usepackage{mathrsfs}
\usepackage{graphicx}
\usepackage{amsmath}
\usepackage{amssymb}
\usepackage{amsfonts}
\usepackage{color}
\usepackage{CJK}
\usepackage{leftidx}

\def\bra#1{\left\langle{#1}\right|}
\def\ket#1{\left|{#1}\right\rangle}
\def\braket#1#2{\left\langle{{#1}}\mathrel{\left|{\vphantom{{#1}{#2}}}\right.\kern-\nulldelimiterspace}{{#2}}\right\rangle}

\begin{document}


\title{Quantum Vector DC Magnetometry via Selective Phase Accumulation}

\author{Min Zhuang}
\affiliation{Institute of Quantum Precision Measurement, State Key Laboratory of Radio Frequency Heterogeneous Integration, Shenzhen University, Shenzhen 518060, China}
\affiliation{College of Physics and Optoelectronic Engineering, Shenzhen University, Shenzhen 518060, China}

\author{Sijie Chen}
\affiliation{Laboratory of Quantum Engineering and Quantum Metrology, School of Physics and Astronomy, Sun Yat-Sen University (Zhuhai Campus), Zhuhai 519082, China}

\author{Jiahao Huang}
\altaffiliation{Email: hjiahao@mail2.sysu.edu.cn}
\affiliation{Laboratory of Quantum Engineering and Quantum Metrology, School of Physics and Astronomy, Sun Yat-Sen University (Zhuhai Campus), Zhuhai 519082, China}

\author{Chaohong Lee}
\altaffiliation{Email: chleecn@szu.edu.cn}
\affiliation{Institute of Quantum Precision Measurement, State Key Laboratory of Radio Frequency Heterogeneous Integration, Shenzhen University, Shenzhen 518060, China}
\affiliation{College of Physics and Optoelectronic Engineering, Shenzhen University, Shenzhen 518060, China}
\affiliation{Quantum Science Center of Guangdong-Hongkong-Macao Greater Bay Area (Guangdong), Shenzhen 518045, China}

\begin{abstract}
  Precision measurement of magnetic fields is an important goal for fundamental science and practical sensing technology.
  Sensitive detection of a vector magnetic field is a crucial issue in quantum magnetometry, it remains a challenge to estimate a vector DC magnetic field with high efficiency and high precision.
  Here, we propose a general protocol for quantum vector DC magnetometry via selective phase accumulation of both non-entangled and entangled quantum probes.
  Based upon the Ramsey interferometry, our protocol may achieve selective phase accumulation of only one magnetic field component by inserting well-designed pulse sequence.
  In the parallel scheme, three parallel quantum interferometries are utilized to estimate three magnetic field components independently.
  In the sequential scheme, by applying a pulse sequence along different directions, three magnetic field components can be estimated simultaneously via only one quantum interferometry.
  In particular, if the input state is an entangled state such as the Greenberger-Horne-Zeilinger state, the measurement precisions of all three components may approach the Heisenberg limit.
  Our study not only develops a general protocol for measuring vector magnetic fields via quantum probes, but also provides a feasible way to achieve Heisenberg-limited multi-parameter estimation via many-body quantum entanglement.
\end{abstract}
\date{\today}

\maketitle

\maketitle
\noindent
\section{Introduction\label{Sec1}}
Precise measurement of a magnetic field is one of the important goals for metrology and sensing.
It is of broad applications in many fields ranging from fundamental physics~\cite{RMP89035002,Helstrom,PRL72,PRL96010401,PRL751879,PRL103081602,Quantum6859,NatureNanotech2301,PNAS10716016}, material science ~\cite{MaterToday14258,Science303}, and geographic metrology to biomedical sensing~\cite{PhysicsReports6150370,SciRep629638,PNAS629638}, fundamental symmetries of nature ~\cite{NaturePhys3227} and dark-matter detection~\cite{NatPhys171402,PRXQUANTUM020101}.
%
%
%
Most of studies focus on scalar magnetic field, which only measure the magnetic field magnitude.
%
Utilizing the well-developed Ramsey techniques, the scalar magnetic field can be detected with ultra-high sensitivity.
However, in practical scenarios, a wide range of magnetometry applications require sensing a vector magnetic field~\cite{NatureNanotech2301,PNAS10716016,PhysRevApplied2018,IMSavukov2005,HXing2016,LRondin2012}, such as magnetic navigation, magnetic-anomaly detection, and biomedical magnetic field detection.
Most vector magnetometers measure the vector field along a specific direction and the determination of all three components requires multiple sensors align along different directions.
%
Therefore, it remains a challenge to employ quantum resources to estimate all three components of a vector DC magnetic field simultaneously~\cite{PRL120080501,MGessner2018,Zhuang2018,Ragy2016}.

Many-body quantum entanglement is a useful resource to improve the measurement precision over individual particles~\cite{Science306,Science316,Nature450,NatPhotonics5,Lucke2011}.
For $N$ individual particles, the measurement precision scales as the standard quantum limit (SQL) (i.e., $\propto 1/\sqrt{N}$).
The SQL can be surpassed by using quantum entanglement.
Especially, by inputting the Greenberger-Horne-Zeilinger (GHZ) states, the measurement precision can be improved to the Heisenberg limit (i.e., $\propto 1/N$)~\cite{RevModPhys90035005,NatPhotonics5222,AnnuRevColdAtomsMolecule,SciRep517894,PRA54R4650,JJBollinger1996}.
Recently, entanglement-enhanced vector DC magnetic field measurement have been proposed~\cite{PRL111070403,PRL116030801,PRL125020501}.
It is indicated that, by employing a mixture of the GHZ-type states in the three directions, the ultimate precisions of all three components can asymptotically approach the Heisenberg limit~\cite{PRL111070403,PRL116030801,PRL125020501}.
These ultimate precisions can achieve via a specific set of positive operator valued measurements (POVM).
However, both the generation of the special mixture state and the realization of the special set of POVM are difficult to be realized with current experimental techniques~\cite{PRL111070403,PRL116030801,PRL125020501}.
Therefore, it is important to find an actual observable or a practical measurement process to achieve the Heisenberg-limited measurement of a vector DC magnetic field.
%

Originated for protecting qubits from decoherence, dynamical decoupling (DD) method becomes one of the well-known quantum control techniques~\cite{Science33060,PRL105053201,SChoi2017,Biercuk2009,Hirose2012,JRMaze2008,GBalasubramanian2008,PRL106080802,LJiang2011,PCMaurer2012,ILovchinsky2016} and has been extensively used to improve the signal-to-noise ratio, from single-particle systems~\cite{Science356837,PRL106080802,Nature473,NatCommun814157,Science356832} to many-body systems~\cite{PRXQUANTUM2040317,PRApplied13044049,PRX10031003}.
%
%
%
%
It is natural to ask: (i) Can one realize a quantum vector DC magnetometry with DD techniques? and
 %
%
(ii) Can the corresponding measurement precisions surpass the SQL or even attain the Heisenberg limit via many-body quantum entanglement?

\begin{figure*}[!htp]
 \includegraphics[width=2\columnwidth]{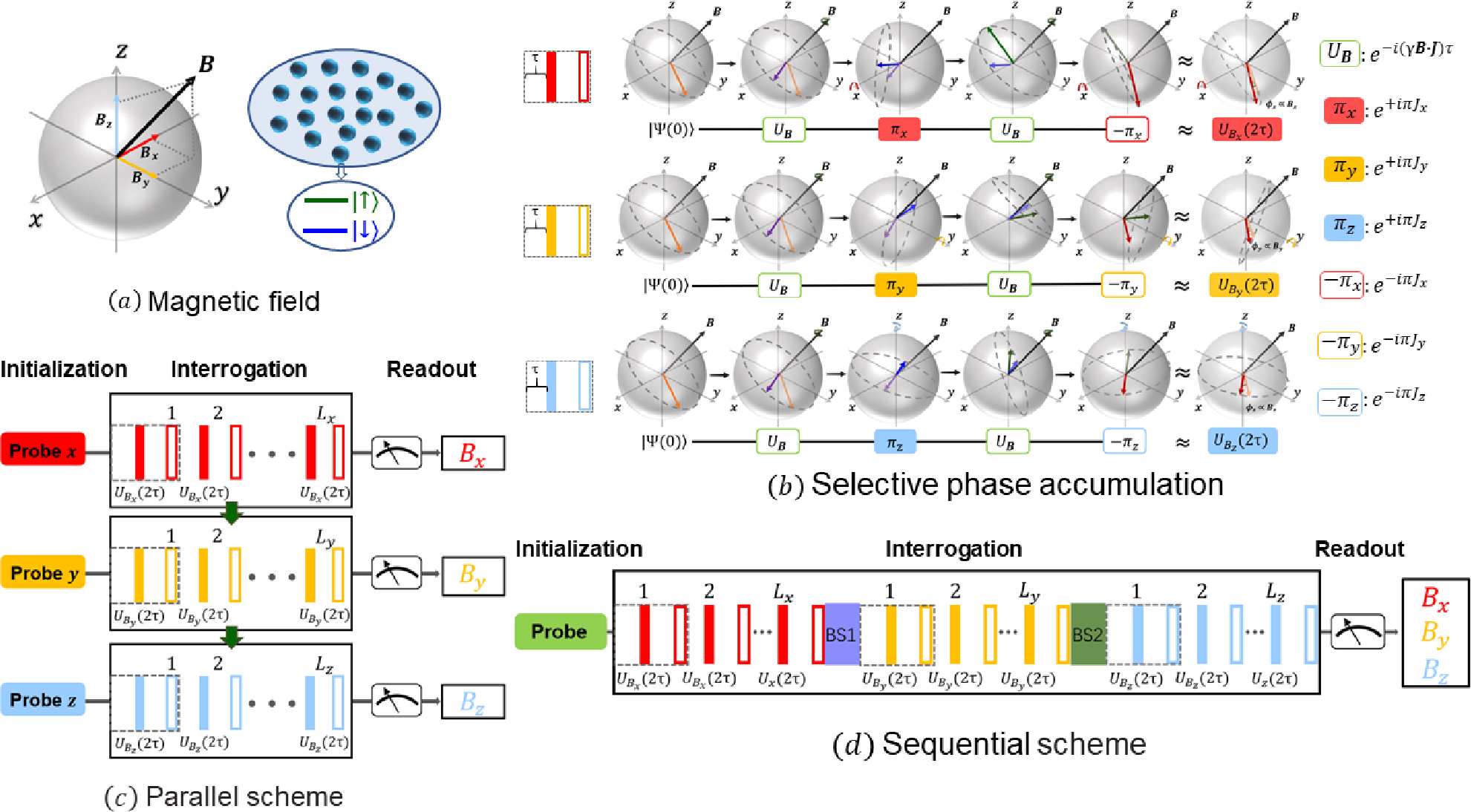}
  \caption{\label{Fig1}
  \textbf{General protocol for measuring a vector DC magnetic field.}
  (a) The interaction between a vector DC magnetic field and an ensemble of two-level particles. %
  (b) The illustration of selective phase accumulation on the Bloch sphere. Left insets: the rapid multi-$\pi_{\alpha}$-pulse during dynamical evolution.
  (c) The parallel scheme for measuring a vector DC magnetic field.
  %
  %
  In the initialization stage, a suitable probe $\alpha$ is prepared for measuring the component ${B_{\alpha}}$, where $\alpha=x,y,z$.
  Then, the probe $\alpha$ undergoes an interrogation stage for signal accumulation, respectively.
  %
  %
  In the readout stage, one can use a practical measurement process to extract the information of the parameter $B_\alpha$.
  (d) The sequential scheme for measuring a vector DC magnetic field.
  In the initialization stage, a suitable probe is prepared.
  Then, the probe undergoes an interrogation stage for signal accumulation.
  The interrogation stage is divided into three signal accumulation processes and linked via two beam splitters.
  In each signal accumulation processes, only one component of the vector magnetic field will contribute to the phase accumulation.
  In the readout stage, one can use a practical measurement process to extract the information of the three components ${B_{\alpha}}$.
  }
\end{figure*}

In this work, we study the estimation of a vector DC magnetic field for both non-entangled and entangled quantum probes via employing multi-pulse quantum interferometry.
The quantum interferometry is implemented by combining Ramsey interferometry and well-designed pulse sequence.
In signal accumulation processes, by applying suitable multiple rapid $\pi$ pulses, only one selected component of the vector magnetic field will accumulate a relative phase between the two levels.
We first consider a parallel scheme with three quantum interferometry, and each is used to selectively estimate one component of the vector DC magnetic field.
For many-body systems with entangled particles, the measurement precisions of the vector field can approach the Heisenberg limit.
However, we need to perform three independent interferometry experiments in such a parallel scheme.
%
%
Thus, to save experimental resources, we propose a sequential scheme to simultaneously estimate all three components of a vector magnetic field in a single interferometry experiment.
The sequential scheme combines Ramsey interferometry with a well-designed $\pi$ pulse-sequence along different directions, and the interrogation is divided into three signal accumulation processes linked with two unitary operations.
%
%
%
%
Moreover, inputting a GHZ state and applying suitable interaction-based operations in both interrogation and readout stages, the measurement precisions of all three components can approach to the Heisenberg limit.
Our scheme may provide a feasible way for measuring a vector DC magnetic field at the Heisenberg limit.
\begin{figure*}[!htp]
 \includegraphics[width=2\columnwidth]{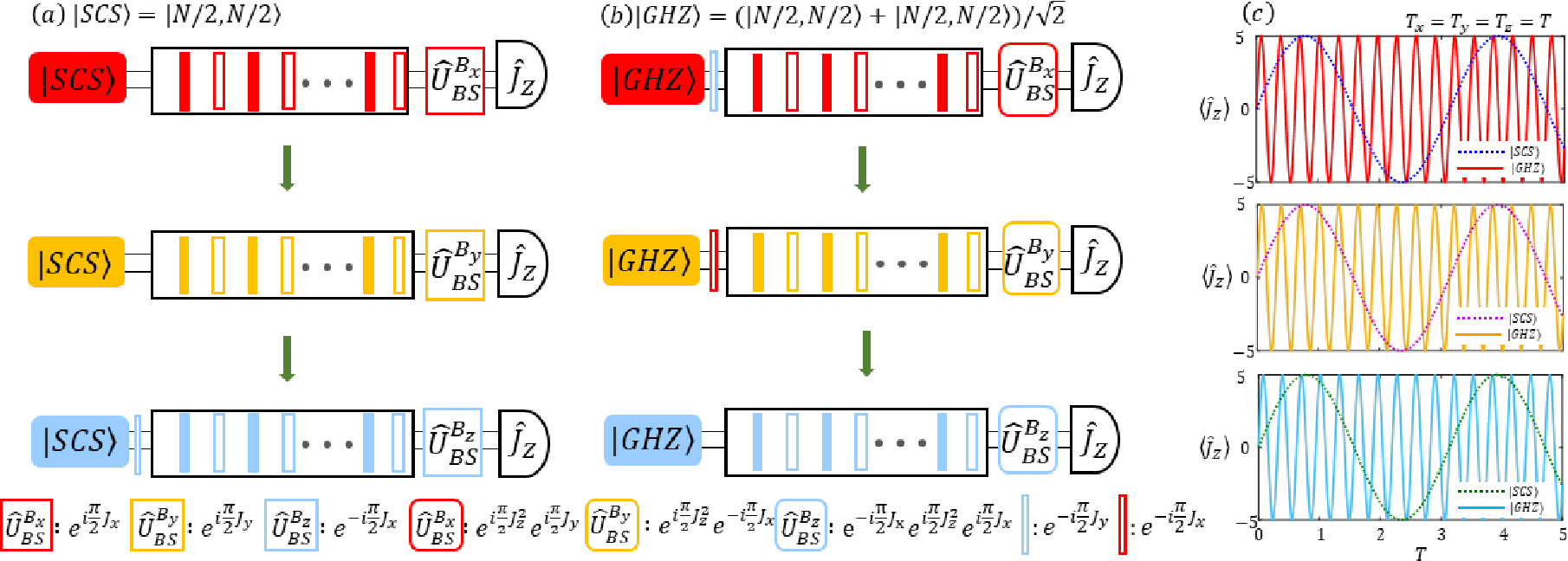}
  \caption{\label{Fig2}
  \textbf{Parallel scheme for measuring a vector DC magnetic field.}
  (a) The parallel scheme with individual particles.
  (b) The parallel scheme with entangled particles.
  (c) The half-population difference versus the interrogation time for the three quantum interferometries with individual and entangled particles.
  Using entangled particles in GHZ state, the oscillation of the half-population difference $\langle J_{z}\rangle$ sensitively depends on the particle number $N$.
  Here, $B_{x}=B_{y}=B_{z}=2$ and $N=10$.}
\end{figure*}

\section{Selective phase accumulation\label{Sec2}}
We consider a probe of $N$ two-level particles coupled to an external vector DC magnetic field ${{\textbf{B}}}= B_{x}{\hat{\textbf{x}}} +B_{y}{\hat{\textbf{y}}}+B_{z}{\hat{\textbf{z}}}$ with $B_{x,y,z}$ standing for its three components, see Fig~\ref{Fig1}~(a).
Potential probes include Bose condensed atoms~\cite{Science345424,Science355620,PNAS1156381,PRL111143001,PRL113103004,Vengalattore2007,CLee2012,BLu2019,Pawlowski2013}, trapped ions~\cite{NaturePhys5,Science3521297,Science3641163,Science3041476,PRL95060502,TMonz2011}, nitrogen-vacancy centers~\cite{NanoLetters215143,NanoLett208267,NanoLett202980,NaturePhys4810}, doped spins in semiconductors~\cite{PRL107166802} and so on.
The two levels are chosen as two magnetic levels, and hereafter are respectively labelled as $\ket{\uparrow}$ and $\ket{\downarrow}$.
Introducing the collective spin operators,
$\hat{J}_{x}=\frac{1}{2}(\hat{a}^{\dag}\hat{b}+\hat{a}\hat{b}^{\dag}),\hat{J}_{y}=\frac{1}{2i}(\hat{a}^{\dag}\hat{b}-\hat{a}\hat{b}^{\dag}),
\hat{J}_{z}=\frac{1}{2}(\hat{a}^{\dag}\hat{a}-\hat{b}^{\dag}\hat{b})$,
the coupling between the probe and the target signal ${\textbf{B}}$ can be described by the Hamiltonian
\begin{equation}\label{Eq:HamS}
\hat{H}_{\textbf{B}}=\gamma{\textbf{B}} \cdot \textbf{J}=\gamma \sum_{\alpha=x,y,z}B_{\alpha}\hat{J}_{\alpha},
\end{equation}
and the probe state can be represented in terms of the Dicke basis $\{|J,m\rangle\}$ with $J=\frac{N}{2}$ and $m = -J,-J + 1, ..., J+1, J$.
Here, $\hat{a}$ and $\hat{b}$ respectively denote annihilation operators for particles in $\ket{\uparrow}$ and $\ket{\downarrow}$, and $\gamma$ is the gyromagnetic ratio.
Our goal is to measure $B_{x}$, $B_{y}$ and $B_{z}$.
Since the generators along different directions are non-commutative, the optimal measurements and probe states for the three components are generally difficult to find.
For single-particle probes, one can utilize an optimal sequential feedback scheme to measure the three components~\cite{PRL117160801,PRL123040501,SciAdv72986}.
However, this scheme requires feedback and it is hard to apply into multi-particle probes.

Inspired by the DD method~\cite{Science33060,Science33060,PRL105053201}, one can apply a suitable sequence of rapid $\pm \pi_{\alpha}$-pulses $(\alpha=x,y,z)$, which are denoted as $e^{\pm i\pi\hat{J}_{\alpha}}$, to selectively retain a desired component $B_{\alpha}$ during time-evolution.
%
%
To analytically show the principle, we assume every $\pm \pi_{\alpha}$ pulse is sharp enough and its time duration can be neglected.
%
%
%
%
In the unit of $\hbar=1$, the time-evolution of probe after a small time interval $2\tau$ is given as
\begin{equation}\label{Eq1}
\ket{\Psi(2\tau)}=e^{-i\pi\hat{J}_{\alpha}}e^{-i\hat{H}_{\textbf{B}}\tau} e^{i\pi\hat{J}_{\alpha}}e^{-i\hat{H}_{\textbf{B}}\tau}\ket{\Psi(0)}
\end{equation}
with the initial state $\ket{\Psi(0)}$ and two neighboring alternating $\pm \pi_{{\alpha}}$-pulses $e^{\pm i \pi\hat{J}_{{\alpha}}}$.
When $\tau$ is sufficiently small, we have $e^{-i\hat{H}_{\textbf{B}}\tau}\approx 1-i\hat{H}_{\textbf{B}}\tau$.
Using the commutation relation of collective spin operators and ignoring the high-order terms (see Appendix A for more details), we have
\begin{equation}\label{Eq2}
\ket{\Psi(2\tau)}\approx e^{-2iB_{{\alpha}}\hat{J}_{{\alpha}}{\tau}}\ket{\Psi(0)}.
\end{equation}
In this way, the probe state $\ket{\Psi(2\tau)}$ can just contain the information of $B_{{\alpha}}$, as shown in Fig~\ref{Fig1}~(b).
According to Eq.~(\ref{Eq2}), with rapid multi-$\pi_{\alpha}$-pulse sequence, the probe state $\ket{\Psi(T_{\alpha})}$ at the time $T_{\alpha}=2L_{\alpha}\tau$ reads as
\begin{eqnarray}\label{Eq3}
\ket{\Psi(T_{\alpha})}&=&\left[e^{-i\pi\hat{J}_{{\alpha}}}e^{-i\hat{H}_{\textbf{B}}\tau} e^{i\pi\hat{J}_{{\alpha}}}e^{-i\hat{H}_{\textbf{B}}\tau}\right]^{L_{\alpha}}\ket{\Psi(0)}\nonumber \\
&\approx& e^{-iB_{{\alpha}}\hat{J}_{{\alpha}}T_{\alpha}}\ket{\Psi(0)}\nonumber \\
\end{eqnarray}
%
%
%
The effective static Hamiltonian under rapid multi-$\pi_{\alpha}$-pulses can be written as $\hat{H}_{B_{\alpha}}^{\text{eff}} = \gamma B_{\alpha} \hat J_\alpha$, see Fig~\ref{Fig1}~(b).

Thus, using rapid multi-$\pi$-pulse along different directions $\alpha=x,y,z$ can separately measure different components $B_{{\alpha}}$ of the vector DC magnetic field.
In our consideration, we find the effective Hamiltonian keeps valid when $\tau\leq 0.002T_{\alpha}$ and this method is feasible with currently available experiment techniques, see Appendix A for details.
%
Under ideal situation, one also can apply rapid identical $\pi_{\alpha}$-pulses to measure $B_{\alpha}$.
However, one need take into account the imperfections of pulses in practical experiments.
One of the most common imperfection of pulse is the rotation angle error.
After some algebra, we find our method with alternating $\pm\pi_{\alpha}$-pulses is more robust to rotation angle errors, see Appendix B for more details.

Based on the selective phase accumulation with rapid multi-$\pi$-pulse, we propose a parallel scheme and a sequential scheme to detect the vector DC field via quantum probes.
According to the parameter quantum estimation theory, the precision of the parameter is constrained by the quantum Cram\'{e}r-Rao bound (QCRB)~\cite{Helstrom,Helstrom1967,Paris2009,AdvPhysX2016},
\begin{eqnarray}\label{Eq5}
  \Delta B_{\alpha} \geq \Delta \mu_\textrm{QCRB}\equiv \frac{1}{\sqrt{\eta F_{Q}^{B_{\alpha}}}},  \quad\alpha=x,y,z,
\end{eqnarray}
with $F_{Q}^{B_{\alpha}}$ is the quantum Fisher information (QFI)(see Appendix C for more details.
In further, we will show that the measurement precision of the three components can reach the Heisenberg limit by using quantum entanglement.

%
%
%
%

\noindent
\section{Parallel scheme\label{Sec3}}
The parallel scheme contains three quantum interferometries, and each interferometry is used to measure one component of the vector DC magnetic field.
As shown in Fig.~\ref{Fig1}~(c), each  quantum interferometry can be divided into three stages: (i) initialization, (ii) interrogation, and (iii) readout.
%
In the initialization stage, a suitable probe ${\alpha}$ is prepared for measuring the component $B_{\alpha}$ .
%
Then, the probe undergoes an interrogation stage for signal accumulation.
The signal accumulation processes of the three different quantum interferometries can be described via the effective static Hamiltonian $\hat{H}_{B_{\alpha}}^{\text{eff}}$ respectively.
According to Eq.~(\ref{Eq3}), only the component of $B_{\alpha}$  will give rise to an accumulated phase $\phi_{\alpha}=B_{\alpha} T_{\alpha}$.
%
%
In the readout stage, one can use a practical measurement process to extract the information of the parameter $B_\alpha$.
Here, we consider a unitary operation for recombination and half-population difference measurement is performed to extract the information of $B_{\alpha}$.
Based upon the parallel scheme, we investigate the measurement precisions with individual and entangled particles.
\subsection{Individual particles}\label{Sec31}
First we discuss the parallel scheme with individual particles.
For individual particles without any entanglement, suppose all particles are prepared in a spin coherent state (SCS) $|\textrm{SCS}\rangle=\ket{N/2,N/2}$, as shown in Fig.~\ref{Fig2}~(a).
For the first quantum interferometer, the final state before the half-population difference measurement is $|\Psi_{\text{final}}^{B_{x}}\rangle=e^{i\frac{\pi}{2}J_{x}}e^{-i\hat{H}_{B_{x}}^{\text{eff}}T_{x}}|\textrm{SCS}\rangle$
%
%
%
%
For the second quantum interferometer, the final state is $|\Psi_{\text{final}}^{B_{y}}\rangle=e^{i\frac{\pi}{2}J_{y}}e^{-i\hat{H}_{B_{y}}^{\text{eff}}T_{y}}|\textrm{SCS}\rangle.$
%
%
For the last quantum interferometer, the final state is $|\Psi_{\text{final}}^{B_{z}}\rangle=e^{-i\frac{\pi}{2}J_{x}}e^{-i\hat{H}_{B_{z}}^{\text{eff}}T_{z}}e^{-i\frac{\pi}{2}J_{y}}|\textrm{SCS}\rangle$.
%
%
%
%
%
%

%
Within the paraller scheme, the quantum Fisher information (QFI) of the three parameters are
\begin{equation}\label{QFI1}
F_{B_{\alpha}}^{\textrm{Q}}=NT_{\alpha}^2, \quad \alpha=x,y,z.
\end{equation}
%
Thus, the ultimate precision bounds with individual particles can just attain the SQL (i.e. $\Delta B_{\alpha}^\textrm{Q} \propto 1/\sqrt{N}$).
%
In further, we consider the measurement precision via measuring half-population difference.
The expectations of final half-population difference for the three quantum interferometries can be written as(see Appendix D for more details)
\begin{eqnarray}\label{Av_Jz}
\langle J_{z} \rangle_{\text{f}}^{B_{\alpha}}=\frac{N}{2}\sin(\phi_{\alpha}),\quad \alpha=x,y,z.
\end{eqnarray}
According to Eq.~\eqref{Av_Jz} one can obtain the information of parameters $B_\alpha$ from the half-population difference.
%
%
Moreover, according to the error propagation formula, we analytically obtain
\begin{eqnarray}
\Delta B_{\alpha}={1}/{\sqrt{N} T_{\alpha}}, \quad \alpha=x,y,z.
\end{eqnarray}
%
\subsection{Entangled particles}\label{Sec32}
In this subsection, we discuss the parallel scheme with entangled particles and show how to realize the Heisenberg-limited measurement of the vector DC magnetic field.
Many-body entanglement is a useful quantum resource to enhance the measurement precision.
In single-parameter estimation, by employing GHZ state as the input state, the measurement precision can be improved to the Heisenberg limit.
Meanwhile, interaction-based readout is a powerful technique for achieving the Heisenberg limit via a GHZ state without single-particle-resolved detection~\cite{PRL116053601,PRL116090801,PRA98012129,PRL119193601,Mirkhalaf2018},
and is now feasible in experiments~\cite{Science3641163,PRL117013001,Science3521552}.
Here, we input the GHZ state $|\textrm{GHZ}\rangle=(\ket{N/2,N/2}+\ket{N/2,-N/2})/\sqrt{2}$ to estimate the vector DC magnetic field, as shown in Fig.~\ref{Fig2}~(b).
For the first quantum interferometer, the final state before measuring the half-population difference is $|\Psi_{\text{final}}^{B_{x}}\rangle=e^{-i\frac{\pi}{2}J_{z}^2}e^{i\frac{\pi}{2}J_{y}}e^{-i\hat{H}_{B_{x}}^{\text{eff}}T_{x}}e^{-i\frac{\pi}{2}J_{y}}|\textrm{GHZ}\rangle$.
%
%
For the second quantum interferometer, the final state is $|\Psi_{\text{final}}^{B_{y}}\rangle=e^{i\frac{\pi}{2}J_{z}^2}e^{-i\frac{\pi}{2}J_{x}}e^{-i\hat{H}_{B_{y}}^{\text{eff}}T_{y}} e^{-i\frac{\pi}{2}J_{x}}|\textrm{GHZ}\rangle$.
%
%
For the last quantum interferometer, the final state is $|\Psi_{\text{final}}^{B_{z}}\rangle=e^{-i\frac{\pi}{2}J_{x}}e^{i\frac{\pi}{2}J_{z}^2}e^{i\frac{\pi}{2}J_{x}}e^{-i\hat{H}_{B_{z}}^{\text{eff}}T_{z}}|\textrm{GHZ}\rangle$.
%
Inputing the GHZ state, the QFI for all three parameters are 
%
\begin{eqnarray}\label{QFI2_SCS}
F_{B_{\alpha}}^{\textrm{Q}}=N^2T_{\alpha}^2, \quad \alpha=x,y,z.
\end{eqnarray}
%
Obviously, the ultimate precision bounds for all three parameters $B_{\alpha}$ now can be improved to the Heisenberg limit via utilizing quantum entanglement (i.e. $\Delta B_{\alpha}^\textrm{Q} \propto 1/{N}$). 

In further, we consider the measurement precision from measuring the half-population difference.
After some algebra (see Appendix E for more details), the expectations of half-population difference are
\begin{eqnarray}\label{Jz_GHZ}
\langle J_{z} \rangle_{\text{f}}^{B_{\alpha}}=\frac{N}{2}\sin(N\phi_{\alpha}), \quad \alpha=x,y,z.
\end{eqnarray}
%
%
%
Similarly, the information of parameters $B_\alpha$ can be inferred from the half-population difference.
Clearly, because of the entanglement, all main frequencies of the bisinusoidal oscillation of half-population difference are proportional to $N = 2J$.
Compared with individual particles, the oscillation frequency with entangled particles is $N$ times higher, as shown in Fig.~\ref{Fig2}~(c).
%
%
%
%
According to the error propagation formula, we have
\begin{equation}\label{DeltaBSCS}
\Delta B_{\alpha}=\frac{{1}}{{{N} T_{\alpha}}}, \quad \alpha=x,y,z.
\end{equation} 
%
%
Thus, the parallel scheme can make the measurement precisions for all three components of a vector DC magnetic field achieve the Heisenberg limit.

\begin{figure*}[!htp]
 \includegraphics[width=1.75\columnwidth]{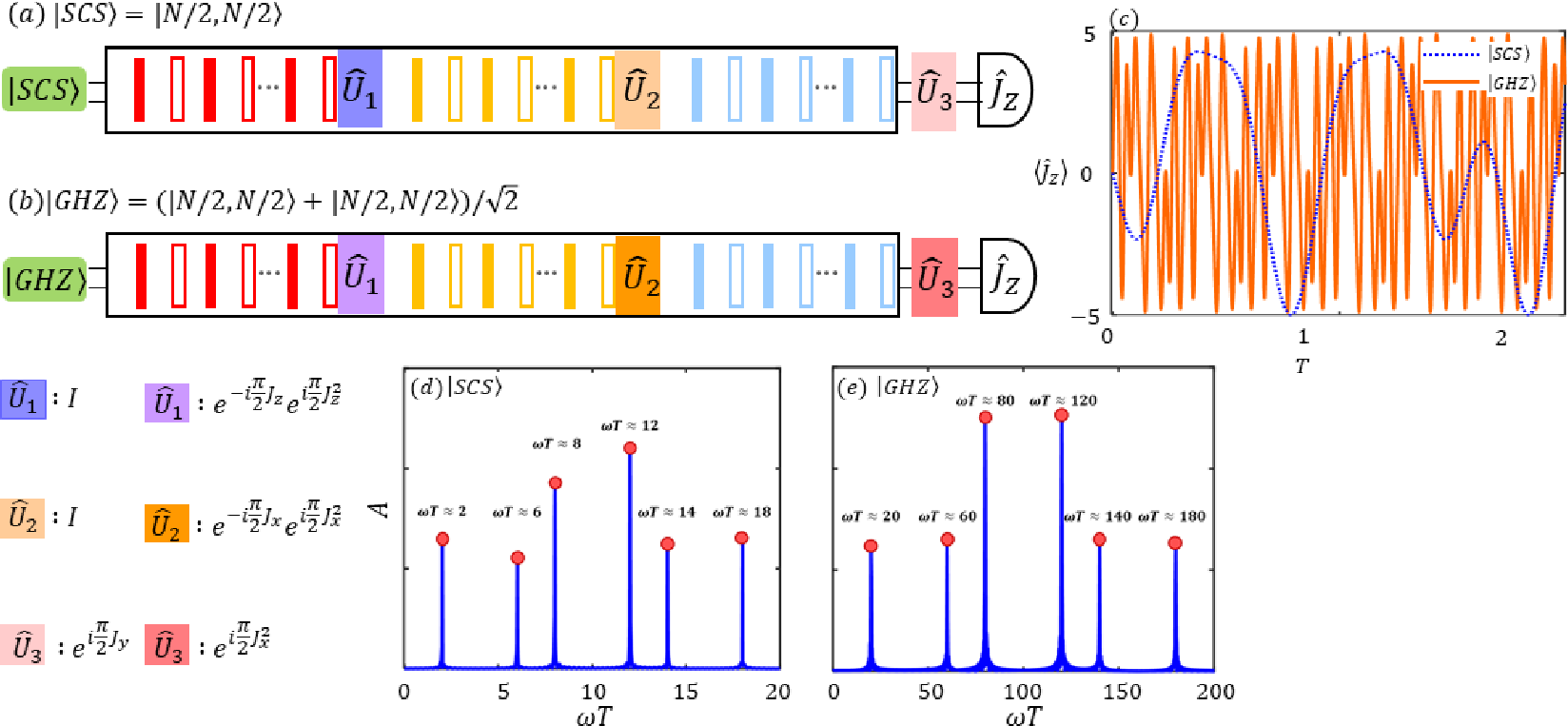}
  \caption{\label{Fig3}
  \textbf{Sequential scheme for measuring a vector DC magnetic field.}
  (a) The sequential scheme with individual particles.
  (b) The sequential scheme with entangled particles.
  (c) The half-population difference versus the interrogation time for individual and entangled particles.
  Inputting a GHZ state, the oscillation of the half-population difference $\langle J_{z}\rangle$ sensitively depends on the particle number $N$.
  %
  %
  %
  (d) FFT spectra of the half-population measurement $\langle J_{z} \rangle_{\text{f}}$ with individual particles in SCS state under $\hat{U}_{1}=\hat{U}_{2}=I$ and $\hat{U}_{3}=e^{-i\frac{\pi}{2}{\hat{J}_{y}}}$.
  The six main oscillation frequencies are very close to $\omega_{1}=B_x+B_y+B_z$, $\omega_{2}=B_x+B_y-B_z$, $\omega_{3}=B_x-B_y+B_z$, $\omega_{4}=B_x-B_y-B_z$, $\omega_{5}=B_x+B_z$ and $\omega_{6}=B_x-B_z$.
  (e) FFT spectra of the half-population measurement $\langle J_{z} \rangle_{\text{f}}$ with entangled particles in GHZ state under $\hat{U}_{1}=e^{-i\frac{\pi}{2}{\hat{J}_{z}}}e^{i\frac{\pi}{2}{\hat{J}_{z}^2}}$, $\hat{U}_{2}=e^{-i\frac{\pi}{2}{\hat{J}_{x}}}e^{i\frac{\pi}{2}{\hat{J}_{x}^2}}$ and $\hat{U}_{3}=e^{i\frac{\pi}{2}{\hat{J}_{x}^2}}$.
  The six main oscillation frequencies are very close to $\omega_{1}=N(B_x+B_y+B_z)$, $\omega_{2}=N(B_x+B_y-B_z)$, $\omega_{3}=N(B_x-B_y+B_z)$, $\omega_{4}=N(B_x-B_y-B_z)$, $\omega_{5}=N(B_x+B_z)$ and $\omega_{6}=N(B_x-B_z)$.
  Here, $B_{x}=10$, $B_{y}=6$, $B_{z}=2$, $N=10$ and $T_{x}=T_{y}=T_{z}=T$....}
\end{figure*}

\section{Sequential scheme\label{Sec4}}
The parallel scheme contains three quantum interferometries and each one needs time for state preparation and readout.
In practical experiments, the time for state preparation and readout need not be negligible.
%
%
Thus, one may save total experimental time if all three components can be measured simultaneously.
%

Here, we illustrate a sequential scheme for simultaneous measurement of all three components of a vector magnetic field.
The sequential scheme includes three stages: (i) initialization, (ii) interrogation, and (iii) readout, see Fig.~\ref{Fig1}~(d).
%
In the initialization stage, a suitable probe is prepared.
Then, the input state undergoes an interrogation stage for signal accumulation.
At this stage, the system state interacts with the magnetic field and three different $\pi_{\alpha}$-pulse sequences are applied rapidly during the interrogation with durations $T_{\alpha}$.
%
%
Under the effect of multi-$\pi$-pulse sequences, the interrogation process can be divided into three signal accumulation processes linked via two beam splitters.
In each signal accumulation process, only one component of the magnetic field will contribute to the phase accumulation.
In order to distinguish the three accumulated phases $\phi_{{\alpha}}$, one needs to rotate the state in the interrogation stage via two beam splitters and they both are unitary operations.
In the readout stage, one can use a practical measurement process to extract the information of all three parameters.
Similarly, we consider use a unitary operation for recombination, and measure the half-population difference to extract the information of the vector magnetic field.
The choices of unitary operations depend on the input state and have influences on the final measurement precisions.
\begin{figure*}[!htp]
 \includegraphics[width=1.75\columnwidth]{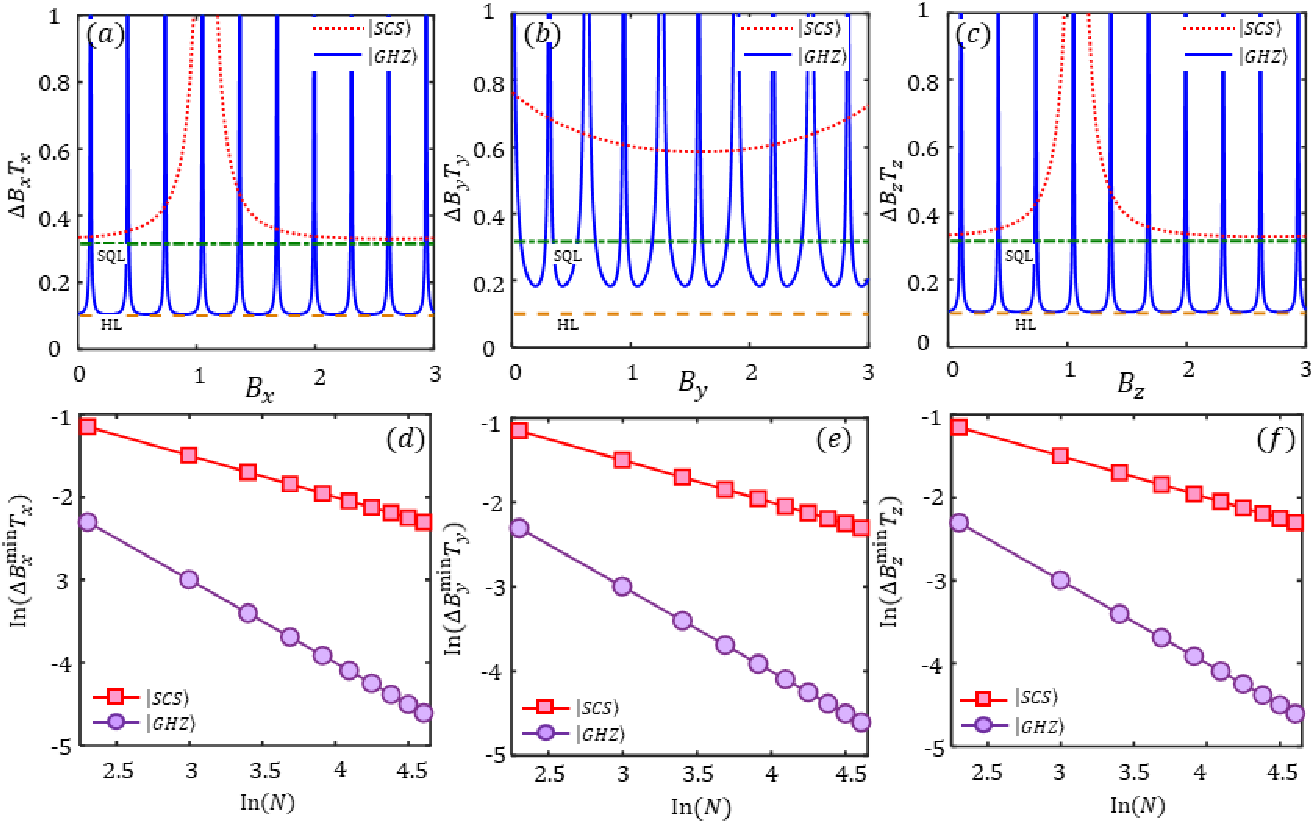}
  \caption{\label{Fig4}
  \textbf{Measurement precisions and their scalings with respect to the particle number.}
  (a) The measurement precision $\Delta B_{x}$ verse $B_{x}$ with $B_{y}=B_{z}=1$.
  (b) The measurement precision $\Delta B_{y}$ verse $B_{y}$ with $B_{x}=B_{z}=1$.
  (c)The measurement precision $\Delta B_{z}$ verse $B_{z}$ with $B_{x}=B_{y}=1$.
  The green and orange dashed lines indicate the standard quantum limit and the Heisenberg limit, respectively.
  Log-log scaling of the optimal measurement precisions for (d) $\Delta B_{x}$, (e) $\Delta B_{y}$ and (f) $\Delta B_{z}$ versus the total particle number $N$.
  The squares and circles correspond to a SCS (with $\hat{U}_{1}=\hat{U}_{2}=I$ and $\hat{U}_{3}=e^{-i\frac{\pi}{2}{\hat{J}_{y}}}$) and a GHZ state (with $\hat{U}_{1}=e^{-i\frac{\pi}{2}{\hat{J}_{z}}}e^{i\frac{\pi}{2}{\hat{J}_{z}^2}}$, $\hat{U}_{2}=e^{-i\frac{\pi}{2}{\hat{J}_{x}}}e^{i\frac{\pi}{2}{\hat{J}_{x}^2}}$ and $\hat{U}_{3}=e^{i\frac{\pi}{2}{\hat{J}_{x}^2}}$), respectively.
  Here $T_{\alpha}=1$ and $N=10$.}
  \end{figure*}
\subsection{Individual particles}\label{Sec41}
First we consider to the sequential scheme with individual particles and assume the probe is prepared in the spin coherent state (SCS) $\ket{\textrm{SCS}}=\ket{N/2,N/2}$.
%
%
Under this situation, one can choose $\hat{U}_{1}=\hat{U}_{2}=I$ and $\hat{U}_{3}=e^{-i\frac{\pi}{2}{\hat{J}_{y}}}$, as shown in Fig.~\ref{Fig3}~(a).
Then, the final state before measuring the half-population difference can be written as
\begin{eqnarray}\label{Evo_CSC}
|\Psi_{\text{final}}\rangle\!=e^{-i\frac{\pi}{2}{\hat{J}_{y}}} e^{-i\hat{H}_{B_{z}}^{\text{eff}}T_{z}} e^{-i\hat{H}_{B_{y}}^{\text{eff}}T_{y}} e^{-i\hat{H}_{B_{x}}^{\text{eff}}T_{x}}\ket{\textrm{SCS}}.\nonumber\\
\end{eqnarray}
%
%
The QFI for the three parameters are 
\begin{eqnarray}\label{QFI1_SCS}
F_{B_{x}}^{\textrm{Q}}=NT_{x}^2,
\end{eqnarray}
\begin{eqnarray}\label{QFI2_SCS}
F_{B_{y}}^{\textrm{Q}}= NT_{y}^2\cos^2(\phi_{x}),
\end{eqnarray}
and
\begin{eqnarray}\label{QFI3_SCS}
F_{B_{z}}^{\textrm{Q}}= NT_{z}^2[1-\cos^2(\phi_{x})\cos^2(\phi_{y})].
\end{eqnarray}
Obviously, the ultimate precision bounds for the three parameters $B_{\alpha}$ can just attain the SQL (i.e. $\Delta B_{\alpha}^\textrm{Q} \propto 1/\sqrt{N}$).

In further, we consider the measurement precision via measuring the half-population difference.
After some algebra, the final half-population difference can be analytically given as (see Appendix F for more details)
\begin{eqnarray}\label{Jz_SCS}
\langle J_{z} \rangle_{\text{f}}&=&\frac{N}{4}[\cos(\phi_{x}+\phi_{z})-\cos(\phi_{x}-\phi_{z})] \nonumber \\ &-&\frac{N}{8}[\sin(\phi_{x}+\phi_{y}+\phi_{z})+\sin(\phi_{x}+\phi_{y}-\phi_{z})]\nonumber \\
&+& \frac{N}{8}[\sin(\phi_{x}-\phi_{y}+\phi_{z})+\sin(\phi_{x}-\phi_{y}-\phi_{z})].\nonumber\\
\end{eqnarray}
%
%
According to Eq.~\eqref{Jz_SCS}, the information of the three parameters $B_{\alpha}$ can be inferred from the multisinusoidal oscillations of the final half-population difference $\langle J_{z} \rangle_{\text{f}}$.
Assuming $T_{\alpha}$ are the same in our calculations, i.e., $T_x=T_y=T_z=T$.
Thus, one can obtain six main oscillation frequencies $\omega_{1}=B_x+B_y+B_z$, $\omega_{2}=B_x+B_y-B_z$, $\omega_{3}=B_x-B_y+B_z$, $\omega_{4}=B_x-B_y-B_z$, $\omega_{5}=B_x+B_z$ and $\omega_{6}=B_x-B_z$ via fast Fourier transform (FFT), and then one can determine the values of $B_x$, $B_y$ and $B_z$.
Firstly, one can determine $B_x$ via the sum of six main oscillation frequencies, i.e, $B_x=({\sum_{i=1}^{6}\omega_{i}})/{6}$.
Then, one can use $|\omega_{i}-B_x|$ to obtain the values of $|B_y+B_z|$, $|B_z|$ and $|B_y-B_z|$.
Ordering the three values from smallest to largest (or largest to smallest), the value of $|B_z|$ is always in the middle of the three values. 
Moreover, the difference between the largest and the smallest values is always equal to $2|B_y|$.
Thus, one can determine the values of $|B_y|$ and $|B_z|$.
Further, one can determine the signs of parameters $B_y$ and $B_z$  from Eq.~\eqref{Jz_SCS} via a fitting procedure.
In addition, the amplitude of oscillation frequencies $\omega_{5}$ and $\omega_{6}$ are higher than others, thus one can determine the value of $|B_z|$ via the values of $\omega_{5}$ and $\omega_{6}$, i.e, $|B_z|=|\omega_{5}-\omega_{6}|/2$.
%
%
%
For individual particles, the oscillations are independent on the total particle number $N$.
In Fig.~\ref{Fig3}~(d), the FFT spectra for $N=10$ with different $B_{x}$, $B_{y}$ and $B_{z}$ are shown.
The numerical results are well agree with our theoretical predictions.
This implies that the values of $B_{x}$, $B_{y}$ and $B_{z}$ can be simultaneously obtained by measuring the final half-population difference.
Further,  one can analytically obtain $\Delta B_{\alpha}$ for individual particles by the the error propagation formula and they
read as (see Appendix F for more details)
\begin{eqnarray}\label{DeltaB1SCS}
\Delta B_{1}\!=\!\frac{1}{\sqrt{N} T_{1}\!}\!\frac{\!\sqrt{1\!-\!G}}{\left| \sin(\phi_{x}) \sin(\phi_{y})\! \cos(\phi_{z})\! -\! \cos(\phi_{x})\! \sin(\phi_{z}) \right|},\nonumber\\
\end{eqnarray}
\begin{eqnarray}\label{DeltaB2SCS}
\Delta B_{2}=\frac{1}{\sqrt{N}T_{2}}\frac{\sqrt{1-G}}{\left|\sin(\phi_{x})\cos(\phi_{y})\cos(\phi_{z}))\right|}\nonumber\\
\end{eqnarray}
and
\begin{eqnarray}\label{DeltaB3SCS}
\Delta  B_{3}=\frac{1}{\sqrt{N}T_{3}}\frac{\sqrt{1-G}}{\left|\cos(\phi_{x})\sin(\phi_{y})\cos(\phi_{z})-\sin(\phi_{x})\sin(\phi_{z})\right|},\nonumber\\
\end{eqnarray}
with
\begin{eqnarray}\label{G}
G=[\cos(\phi_{x})\sin(\phi_{y})\cos(\phi_{z})+\sin(\phi_{x})\sin(\phi_{z})]^2 \nonumber\\
\end{eqnarray}
%
The measurement precisions $\Delta B_{\alpha}$ versus $B_{\alpha}$ are shown in Fig.~\ref{Fig4}~(a)-(c) (red dotted lines).
It is obvious that the measurement precisions $\Delta B_{\alpha}$ follows the SQL scaling (i.e., $\Delta B_{\alpha}\propto1/\sqrt{N}$).
Further, we numerically find the optimal measurement precisions $\Delta B_{\alpha}^{\textrm{min}}$ versus the total particle number $N$, as shown in Fig.~\ref{Fig4}.
According to the fitting results, the optimal log-log measurement precisions for $\textrm{ln}\left(\Delta B_{x}^{\textrm{min}}\right)\approx -0.5\textrm{ln}({N})$ (red squares), $\textrm{ln}\left(\Delta B_{y}^{\textrm{min}}\right)\approx -0.5\textrm{ln}({N})$ (red squares) and $\textrm{ln}\left(\Delta B_{z}^{\textrm{min}}\right)\approx -0.5\textrm{ln}({N})$ (red squares).
Since the input state is not entangled, the optimal measurement precisions for the three parameters just can saturate the SQL.
\subsection{Entangled particles}\label{Sec32}
In this subsection, we discuss the sequential scheme with entangled particles and show how it may realize the Heisenberg-limited measurement.
We choose the GHZ state $|\textrm{GHZ}\rangle=(\ket{N/2,N/2}+\ket{N/2,-N/2})/\sqrt{2}$ as an input state and three suitable interaction-based operations to perform a simultaneous measurement.
The three different interaction-based operations are $\hat{U}_{1}=e^{-i\frac{\pi}{2}{\hat{J}_{z}}}e^{i\frac{\pi}{2}{\hat{J}_{z}^2}}$, $\hat{U}_{2}=e^{-i\frac{\pi}{2}{\hat{J}_{x}}}e^{i\frac{\pi}{2}{\hat{J}_{x}^2}}$ and $\hat{U}_{3}=e^{i\frac{\pi}{2}{\hat{J}_{x}^2}}$, see Fig.~\ref{Fig3}~(b).
After the sequence, the final state can be written as
\begin{eqnarray}\label{Evo_Spin_cat_state3}
\ket{\Psi_{\text{final}}}&&= e^{i\frac{\pi}{2}{\hat{J}_{x}^{2}}} e^{-i\hat{H}_{B_{z}}^{\text{eff}}T_{z}}e^{i\frac{\pi}{2}{\hat{J}_{x}^2}}e^{-i\frac{\pi}{2}{\hat{J}_{x}}}
\nonumber \\&&\times e^{-i\hat{H}_{B_{y}}^{\text{eff}}T_{y}}e^{i\frac{\pi}{2}{\hat{J}_{z}^{2}}}e^{-i\frac{\pi}{2}{\hat{J}_{z}}} e^{-i\hat{H}_{B_{x}}^{\text{eff}}T_{x}}
\ket{\textrm{GHZ}}.
\end{eqnarray}
If $N$ is an even number, the QFI for the three parameters with GHZ are 
\begin{eqnarray}\label{QFI1_SCS}
F_{B_{x}}^{\textrm{Q}}=N^2T_{x}^2,
\end{eqnarray}
\begin{eqnarray}\label{QFI2_SCS}
F_{B_{y}}^{\textrm{Q}}= N^2T_{y}^2\cos^2(\phi_{x})
\end{eqnarray}
and
\begin{eqnarray}\label{QFI3_SCS}
F_{B_{z}}^{\textrm{Q}}= N^2T_{z}^2[1-\cos^2(\phi_{x})\cos^2(\phi_{y})]
\end{eqnarray}
By using entangled particle in GHZ states, the ultimate precision bounds can be improved to the Heisenberg limit(i.e. $\Delta B_{\alpha}^\textrm{Q} \propto 1/{N}$).
%
%
%

Further, after some algebra (see Appendix G for more details), the final half-population difference is
%
\begin{eqnarray}\label{Jz_GHZ}
\langle J_{z} \rangle_{\text{f}}&&=\frac{N}{8}\left[\sin(N(\phi_{x}+\phi_{y}+\phi_{z}))+\sin(N(\phi_{x}-\phi_{y}+\phi_{z}))\right]\nonumber \\
&&-\frac{N}{8}[\sin(N(\phi_{x}+\phi_{y}-\phi_{z}))+\sin(N(\phi_{x}-\phi_{y}-\phi_{z}))]\nonumber \\
&&-(-1)^{J}\frac{N}{4}[\sin(N (\phi_{x}+\phi_{z}))+\sin(N(\phi_{x}-\phi_{z}))]
\end{eqnarray}
%
Due to the entanglement, the main frequencies of multisinusoidal oscillation of $\langle J_{z} \rangle_{\text{f}}$ becomes proportional to $N=2J$, and the oscillation frequency is higher than the one for individual particles, as shown in Fig.~\ref{Fig3}~(c).
In the case of $T_x=T_y=T_z=T$, the FFT of $\langle J_{z} \rangle_{\text{f}}$ explicitly indicates that the six main oscillation frequencies are
$\omega_{1}=N(B_x+B_y+B_z)$, $\omega_{2}=N(B_x+B_y-B_z)$, $\omega_{3}=N(B_x-B_y+B_z)$, $\omega_{4}=N(B_x-B_y-B_z)$, $\omega_{5}=N(B_x+B_z)$ and $\omega_{6}=N(B_x-B_z)$.
Under this situation, one also can distinguish the value of $B_x$, $B_y$, and $B_z$ via the method illustrated for individual particles.
The numerical results perfectly agree with our theoretical predictions, see Fig.~\ref{Fig3}~(b).
It indicates that the values of $B_{x}$, $B_{y}$ and $B_{z}$ can be simultaneously obtained by measuring the half-population difference.
Moreover, the square of half-population difference is independent on the three parameters, i.e., $\langle J_{z}^{2} \rangle_{\text{f}}={J^{2}}=N^2/4$.
According to the error propagation formula, we analytically obtain $\Delta B_{\alpha}$ and they are
\begin{widetext}
\begin{eqnarray}\label{DeltaB1GHZ}
\Delta B_{1}=\frac{1}{{N}T_{x}}\frac{\sqrt{1-[\cos(N\phi_{x})\cos(N\phi_{y})\sin(N\phi_{z})-(-1)^{J}\sin(N\phi_{x})\cos(N\phi_{z})]^2}}
{\left|\sin(N\phi_{x})\cos(N\phi_{y})\sin(N\phi_{z})+(-1)^{J}\cos(N\phi_{x})\cos(N\phi_{z})\right|},
\end{eqnarray}
\begin{eqnarray}\label{DeltaB2GHZ}
\Delta B_{2}=\frac{1}{{N}T_{y}}\frac{\sqrt{1-[\cos(N\phi_{x})\cos(N\phi_{y})\sin(N\phi_{z})-(-1)^{J}\sin(N\phi_{x})\cos(N\phi_{z})]^2}}
{\left|\cos(N\phi_{x})\sin(N\phi_{y})\sin(N\phi_{z})\right|},
\end{eqnarray}
and
\begin{eqnarray}\label{DeltaB3GHZ}
\Delta B_{3}=\frac{1}{{N}T_{z}}\frac{\sqrt{1-[\cos(N\phi_{x})\cos(N\phi_{y})\sin(N\phi_{z})-(-1)^{J}\sin(N\phi_{x})\cos(N\phi_{z})]^2}}
{\left|\cos(N\phi_{x})\cos(N\phi_{y})\cos(N\phi_{z})+(-1)^{J}\sin(N\phi_{x})\sin(N\phi_{z})\right|}.
\end{eqnarray}
\end{widetext}

According to Eq.~\eqref{DeltaB1GHZ} $\sim$ Eq.~\eqref{DeltaB3GHZ}, the measurement precisions $\Delta B_{\alpha}$ for individual particles can exhibit the HL scaling(i.e.,$\Delta B_{\alpha}^\textrm{Q} \propto 1/{N}$).
In Fig.~\ref{Fig4}~(a)-(c), we show the measurement precision $\Delta B_{\alpha}$ versus $B_{\alpha}$ (solid blue lines).
%
%
%
Furthermore, we numerically find the minimum measurement precisions of $\Delta B_{\alpha}^{\textrm{min}}$ versus the particle number. 
As shown in Fig.~\ref{Fig4}, the optimal log-log measurement precisions $\Delta B_{x}^{\textrm{min}}$ (purple circles), $\Delta B_{y}^{\textrm{min}}$ (purple circles) and $\Delta B_{z}^{\textrm{min}}$ (purple circles) are $\textrm{ln}\left(\Delta B_{x}^{\textrm{min}}\right)\approx -\textrm{ln}({N})$ , $\textrm{ln}\left(\Delta B_{y}^{\textrm{min}}\right)\approx -\textrm{ln}({N})$ and $\textrm{ln}\left(\Delta B_{z}^{\textrm{min}}\right)\approx -\textrm{ln}({N})$.
This indicates that our scheme enables to measure all three components of a vector DC magnetic field simultaneously and the measurement precisions can approach the Heisenberg limit.

\section{summary and discussions\label{Sec4}}
With state-of-the-art techniques, we study the vector DC magnetic field estimation via employing multi-pulse quantum interferometry.
%
The interferometry is implemented by combining the Ramsey interferometry and different rapid-pulse sequences, which does not need multiple sensors along different directions.
In signal accumulation processes, by applying suitable multiple rapid $\pi_{\alpha}$-pulses, only one component $B_{\alpha}$ of the magnetic field will give rise to an phase accumulation.
In the parallel scheme, each many-body quantum interferometry can estimate one component of the vector DC magnetic field.
Further, to save the total experimental time, we propose the sequential scheme.
The sequential scheme combines the Ramsey interferometry with well-designed $\pi$-pulse sequences along different directions, and can estimate all three components of the vector DC magnetic field simultaneously via the population measurement.
The interrogation process in the sequential scheme is divided into three signal accumulation processes linked by two unitary operations.
For both schemes, we analytically obtain the measurement precisions of the three components with individual and entangled particles.
We find that, by using entangled particles in GHZ state and applying suitable interaction-based operations, the measurement precisions can achieve the Heisenberg-limited scaling.

To realize our quantum magnetometry in experiments, one has to combine the Ramsey interferometry with multiple rapid $\pi$-pulses in the interrogation stage.
The precise implementation of $\pi$-pulses is a mature technology in quantum control, especially for non-entangled single-particle systems.
Meanwhile, our schemes are robust against imperfect $\pi$-pulses with rotation angle errors.

Furthermore, to achieve an entanglement-enhanced magnetometry with Heisenberg-limited measurement in experiments, the preparation of the desired GHZ state and the implementation of an interaction-based readout are important.
Owing to the well-developed techniques in quantum control, various multi-particle entangled states have been generated in several systems, including nitrogen-vacancy defect centers~\cite{PRL93130501}, Bose condensed atoms~\cite{Science345424,PRL117013001,Nature4641165,Nature4641170,Science355620,PNAS1156381}, ultracold trapped ions~\cite{Science3521297,Science3521552}, and solid-state spin systems.
In particular, the GHZ state can be generated in an ensemble of Bose condensed atoms occupying two hyperfine levels via dynamical evolution~\cite{Nature4641165,Nature4641170,PRL90030402,PRL107013601,AP1900471} or adiabatic evolution~\cite{SciRep517894,PRL97150402,PRL102070401,PRA97032116,PRL111180401}under a one-axis twisting Hamiltonian $\hat{H}_\textrm{twist}=\chi\hat{J}_{z}^2+\Omega \hat{J}_{x}$.
The interaction-based operations can also be realized in experiment via modulating the nonlinear inter-particle interaction~\cite{PRL90030402,PRL107013601,Science345424,Science355620}.
The strength and the sign of the nonlinearity $\chi$, determined by the $s$-wave scattering lengths and the spatial overlap between different spin components, can be tuned via applying a spin-dependent force~\cite{Nature4641170,PRL111143001} or the techniques of Feshbach resonance~\cite{Science345424,Nature4641170,PRL113103004}.
%
Our study not only paves a new way for measuring a vector magnetic field with quantum systems, but also provides a feasible method for achieving Heisenberg-limited multiparameter estimation via many-body quantum entanglement.

\section*{Acknowledgements}
\noindent
This work is supported by the National Key Research and Development Program of China (Grant No. 2022YFA1404104), the National Natural Science Foundation of China (Grant No. 12025509), and the Key-Area Research and Development Program of GuangDong Province (Grant No. 2019B030330001).
\setcounter{equation}{0}
\renewcommand{\theequation}{A\arabic{equation}}

\section*{APPENDIX A: Derivation of the effective Hamiltonian\label{SecS1}}
Here, we only show how to derive the effective Hamiltonian $\hat{H}_{B_{x}}^{\text{eff}} = \gamma B_{x} \hat J_x$, as the other two effective Hamiltonians $\hat{H}_{B_{y}}^{\text{eff}}$ and $\hat{H}_{B_{z}}^{\text{eff}}$ can be obtained similarly.
If $\tau$ is sufficiently small, one can assume $e^{-i\hat{H}_{\textbf{B}}\tau}\approx 1-i\hat{H}_{\textbf{B}}\tau$ and thus the probe state $\ket{\Psi(2\tau)}$ can be written as
\begin{eqnarray}\label{EqM1}
\ket{\Psi(2\tau)}&&=e^{-i\pi\hat{J}_{x}}e^{-i\hat{H}_{\textbf{B}}\tau} e^{i\pi\hat{J}_{x}}e^{-i\hat{H}_{\textbf{B}}\tau}\ket{\Psi(0)}\nonumber \\
&&\approx \!  [1\!-\!ie^{-i\pi\hat{J}_{x}}\hat{H}_{\textbf{B}}e^{i\pi\hat{J}_{x}}\tau\!-\!i\hat{H}_{\textbf{B}}\tau]\ket{\Psi(0)}.
\end{eqnarray}
Here, the approximation in Eq.~\eqref{EqM1} is just retain to the first-order term, i.e., $\propto \tau$.
By utilizing the commutation relation of collective spin operators, $[\hat{J}_a, \hat{J}_b]=i\epsilon_{abc}\hat{J}_c$ (with $\epsilon_{abc}$ being the Levi-Civita symbol and $a,b,c=x,y,z$), we have
\begin{eqnarray}\label{EqM2}
e^{-i\pi\hat{J}_{x}}\hat{H}_{\textbf{B}}e^{i\pi\hat{J}_{x}}=B_{x}\hat{J}_{x}-B_{y}\hat{J}_{y}-B_{z}\hat{J}_{z}.
\end{eqnarray}
Submitting Eq.~\eqref{EqM2} into Eq.~\eqref{EqM1}, we have
\begin{eqnarray}\label{EqM3}
|\Psi(2\tau)\rangle\!\approx\!(1\!-\!2iB_{x}\!\hat{J}_{x}\tau)|{\Psi(0)\!}\rangle\!\approx\! e^{\!-\!2iB_{x}\hat{J}_{x}\tau}|{\Psi(0)\!}\rangle.
\end{eqnarray}
Thus the dynamical process under repaid $\pm \pi_{x}$-pulse can be described by the effective static Hamiltonian $\hat{H}_{B_{x}}^{\text{eff}} = \gamma B_{x} \hat J_x$.
In the same way, the dynamical process under repaid $\pm \pi_{y}$-pulse can be described by the effective static Hamiltonian $\hat{H}_{B_{y}}^{\text{eff}} = \gamma B_{y} \hat J_y$,
and the dynamical process under repaid $\pm \pi_{z}$-pulse can be described by the effective static Hamiltonian $\hat{H}_{B_{z}}^{\text{eff}} = \gamma B_{z} \hat J_z$.
The validity of the effective Hamiltonian is important to realize the quantum vector DC magnetometry.
To illustrate the validity of the effective Hamiltonian and give a significant reference to practical experiments, we numerically calculate the fidelity $F_{1}(t)=|\langle \psi(t)|\psi(t)_{\textrm{eff}}\rangle|^2$ between the exact and effective evolved states.
The two evolved states are
\begin{eqnarray}\label{EqM6}
\ket{\Psi(t)}=&&\left[e^{-i\pi\hat{J}_{x}}e^{-i\hat{H}_{\textbf{B}}\tau} e^{i\pi\hat{J}_{x}}e^{-i\hat{H}_{\textbf{B}}\tau}\right]^{L_x}\nonumber\\
\times&&\left[e^{-i\pi\hat{J}_{y}}e^{-i\hat{H}_{\textbf{B}}\tau} e^{i\pi\hat{J}_{y}}e^{-i\hat{H}_{\textbf{B}}\tau}\right]^{L_y}\nonumber\\
\times&&\left[e^{-i\pi\hat{J}_{z}}e^{-i\hat{H}_{\textbf{B}}\tau} e^{i\pi\hat{J}_{z}}e^{-i\hat{H}_{\textbf{B}}\tau}\right]^{L_z}\ket{\Psi(0)}, \nonumber\\
\end{eqnarray}
and
\begin{eqnarray}\label{EqM6}
\ket{\Psi(t)}_{\textrm{eff}}=e^{-i{2L_{x}\tau}\hat{H}_{B_{x}}^{\text{eff}}}e^{-i2L_{y}\tau\hat{H}_{B_{y}}^{\text{eff}}}e^{-i2L_{z}\tau\hat{H}_{B_{z}}^{\text{eff}}}\ket{\Psi(0)}, \nonumber\\
\end{eqnarray}
with $t=2(L_{x}+L_{y}+L_{z})\tau$.
Setting the evolution time $T =6$, $L_{x}=L_{y}=L_{z}=L$ and $N=10$, we numerically find that the two evolved states are almost the same when $\tau/T\leq0.002$, i.e., $F_{1}(t)\approx1$, see Fig.~\ref{FigS1}~(a).

\begin{figure}[!htp]
 \includegraphics[width=\columnwidth]{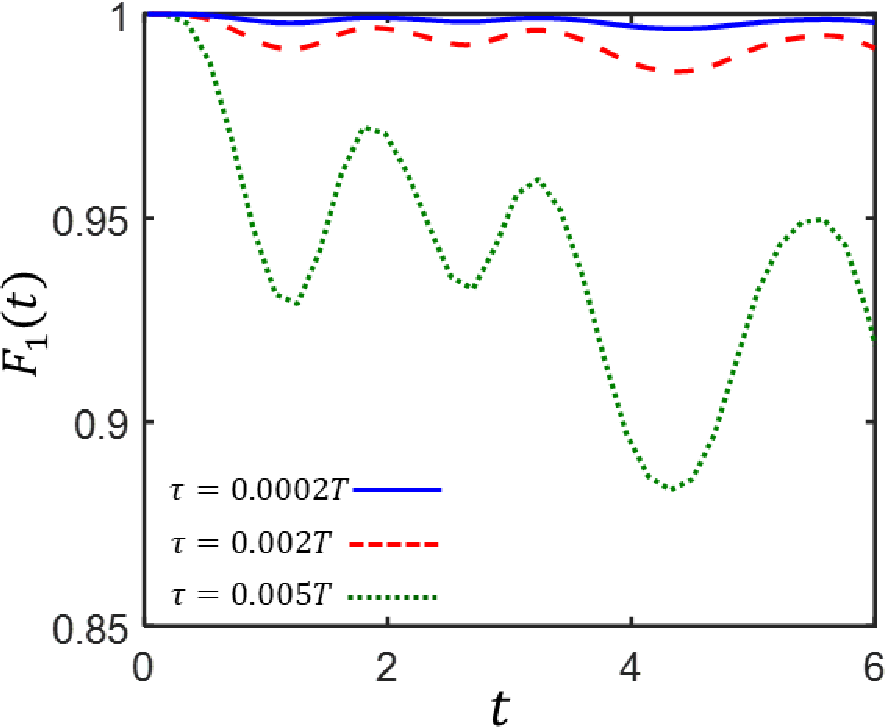}
  \caption{\label{FigS1}
  \textbf{The validity of the effective Hamiltonian.}
  (a) The fidelity $F_{1}(t)$ versus the evolution time $t$ for different pulse spacing $\tau$.
  The solid-blue, dashed-red and dotted-green lines correspond to $\tau=0.0002T$, $0.002T$ and $0.005T$, respectively.
  Here $N=10$, $\tau=10^{-3}$, $B_{x}=4$, $B_{y}=5$, $B_{z}=6$, $T=6$.
  %
  %
  Here $N=10$, $\tau=10^{-3}$, $B_{x}=4$, $B_{y}=5$, $B_{z}=6$.}
\end{figure}

\section*{APPENDIX B: Robustness against rotation angle errors\label{SecS2}}
%
To realize perfect interrogation in our schemes, the pulses should be ideal $\pi$ pulses.
However, due to the instability of the external control fields, one need take into account the imperfections of pulses in practical experiments.
One of the most common imperfection of pulse is the rotation angle error.
For an imperfect $\pi$-pulse with a rotation angle error $\emph{d}\theta$, the pulse becomes $e^{-i(\pi+\emph{d}\theta)\hat{J}_{\alpha}}$.
Here, we take $\pi_{x}$ pulse as an example, thus the corresponding evolved state $\ket{\Psi(2\tau)}$ can be written as
\begin{eqnarray}\label{EqM4}
\ket{\Psi(2\tau)}=e^{-i(\pi+\emph{d}\theta)\hat{J}_{x}}e^{\!-i\hat{H}_{\textbf{B}}\tau} e^{i(\pi+\emph{d}\theta)\hat{J}_{x}}e^{-i\hat{H}_{\textbf{B}}\tau}|{\Psi(0)}\rangle.\nonumber \\
\end{eqnarray}
Further, by using the commutation relation of collective spin operators and just retaining the first-order term, we have
\begin{eqnarray}\label{EqM5}
\ket{\Psi(2\tau)}&&\approx[1-2iB_{x}\hat{J}_{x}]\ket{\Psi(0)}\nonumber \\
&&-i(B_{2}\!-\!\cos(\emph{d}\theta)B_{y}\!-\!\sin(\emph{d}\theta)B_{z})\hat{J}_{y}\ket{\Psi(0)} \nonumber \\
&&-i(B_{3}\!-\!\cos(\emph{d}\theta)B_{z}\!-\!\sin(\emph{d}\theta)B_{y})\hat{J}_{z}\ket{\Psi(0)}.\nonumber \\
\end{eqnarray}
%
%
To illustrate the robustness of our schemes, we assume the rotation angle error $\emph{d}\theta(t)$ is time-dependent and the fluctuation of rotation angle error $\emph{d}\theta(t)$ satisfies $\emph{d}\theta(t)\in [-\eta, \eta]$, where $\eta$ is the maximum fluctuation strength.
We numerically calculate the fidelity $F_2(t)=|_{\eta=0}\langle \Psi(t)|\Psi(t)\rangle_{\eta}|^2$
between the two evolved states $\ket{\Psi(t)}_{\eta=0}$ and $\ket{\Psi(t)}_{\eta}$.
The evolved states $\ket{\Psi(t)}_{\eta}$ can be written as
\begin{small}
\begin{eqnarray}\label{EqM6}
\ket{\Psi(t)}_{\eta}&=&\left[e^{-i(\pi+\emph{d}\theta)\hat{J}_{x}}e^{-i\hat{H}_{\textbf{B}}\tau} e^{i(\pi+\emph{d}\theta)\hat{J}_{x}}e^{-i\hat{H}_{\textbf{B}}\tau}\right]^{L_{x}}\nonumber\\
&\times&\left[e^{-i(\pi+\emph{d}\theta)\hat{J}_{y}}e^{-i\hat{H}_{\textbf{B}}\tau} e^{i(\pi+\emph{d}\theta)\hat{J}_{y}}e^{-i\hat{H}_{\textbf{B}}\tau}\right]^{L_{y}}\nonumber\\
&\times&\left[e^{-i(\pi+\emph{d}\theta)\hat{J}_{z}}e^{-i\hat{H}_{\textbf{B}}\tau} e^{i(\pi+\emph{d}\theta)\hat{J}_{z}}e^{-i\hat{H}_{\textbf{B}}\tau}\right]^{L_{z}}\ket{\Psi(0)}, \nonumber\\
\end{eqnarray}
\end{small}
with $t=2(L_{x}+L_{y}+L_{z})\tau$.
Similarly, setting $L_{x}=L_{y}=L_{z}=L$, we numerically calculate the fidelity $F_2(t)$ under different rotation angle errors with $\eta=0.02\pi$, $\eta=0.06\pi$ and $\eta=0.1\pi$, and after averaging $20$ times, see Fig.~\ref{FigS1}~(b).
Our numerical simulation clearly indicates that, if $\eta\leq0.06\pi$, the two evolved states are almost the same, i.e., $F_{2}(t)\approx1$.
Thus our scheme is robust against rotation angle errors.

For ideal situation without rotation angle errors, one also can apply a suitable sequence of rapid $\pi_{\alpha}$-pulses $(\alpha=x,y,z)$.
However when the rotation angle error exists, the evolved state $\ket{\Psi(2\tau)}$ with $\pi_{x}$ pulses becomes
\begin{eqnarray}\label{EqM6}
|{\Psi(2\tau)}\rangle\!=\!e^{\!-i(\!\pi+\emph{d}\theta)\hat{J}_{x}}\!e^{\!-i\hat{H}_{\textbf{B}}\tau} \!e^{\!-i(\!\pi+\emph{d}\theta)\hat{J}_{x}}\!e^{\!-i\!\hat{H}_{\textbf{B}}\tau}|{\Psi(0)}\rangle. \nonumber\\
\end{eqnarray}
In an explicit form, the state $\ket{\Psi(2\tau)}$ becomes
\begin{small}
\begin{eqnarray}\label{EqM7}
\ket{\Psi(2\tau)}\approx&&[1-2iB_{x}\tau\hat{J}_{x}]e^{-2i(\pi+\emph{d}\theta)\hat{J}_{x}}\ket{\Psi(0)}\nonumber \\
&&-i\left[\left(\cos(2\emph{d}\theta)-\cos(\emph{d}\theta)\right)B_{y}\right]\hat{J}_{y}e^{-2i(\pi+\emph{d}\theta)\hat{J}_{x}}|{\Psi(0)}\rangle \nonumber \\
&&-i\left[\left(\sin(2\emph{d}\theta)-\sin(\emph{d}\theta)\right)B_{z}\right]\hat{J}_{y}e^{-2i(\pi+\emph{d}\theta)\hat{J}_{x}}|{\Psi(0)}\rangle \nonumber \\
&&-i\left[\left(\cos(2\emph{d}\theta)-\cos(\emph{d}\theta)\right)B_{z}\right]\hat{J}_{z}e^{-2i(\pi+\emph{d}\theta)\hat{J}_{x}}|{\Psi(0)}\rangle.\nonumber \\
&&+i\left[\left(\sin(2\emph{d}\theta)+\sin(\emph{d}\theta)\right)B_{y}\right]\hat{J}_{z}e^{-2i(\pi+\emph{d}\theta)\hat{J}_{x}}|{\Psi(0)}\rangle.\nonumber \\
\end{eqnarray}
\end{small}
Comparing Eq.~\eqref{EqM5} with Eq.~\eqref{EqM7}, one can find that the rotation angle error $\emph{d}\theta(t)$ has more negative influences on the evolved state when the two pulses are the same.
Thus the sequence of rapid $\pm \pi_{\alpha}$-pulses is more robust against rotation angle errors than the sequence of rapid $\pi_{\alpha}$-pulses.
\begin{figure}[!htp]
 \includegraphics[width=1\columnwidth]{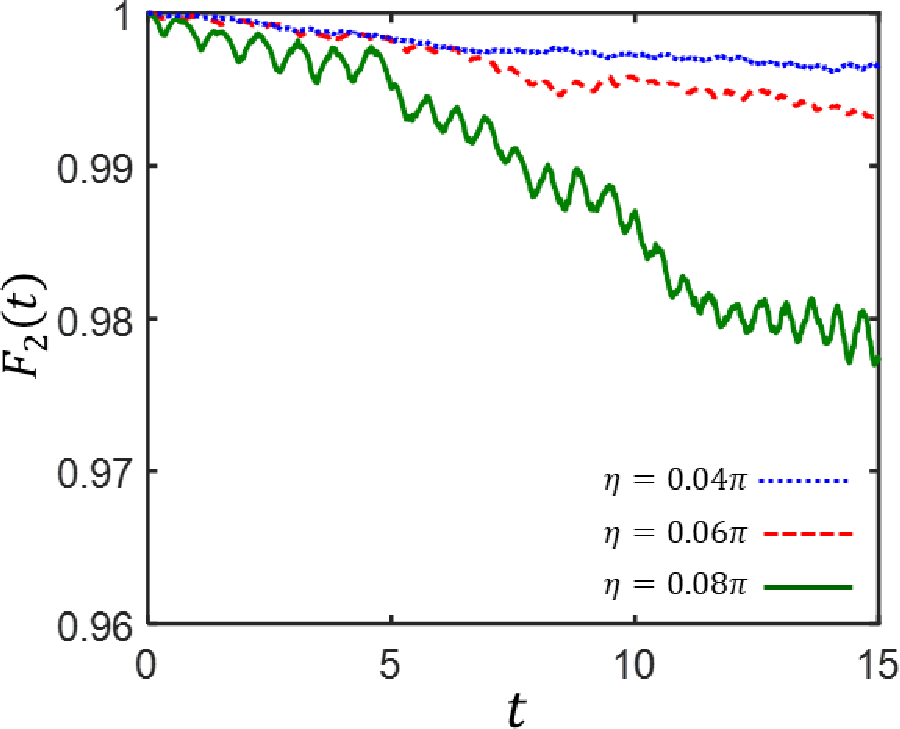}
  \caption{\label{FigS1}
  \textbf{The robustness of our scheme against rotation angle errors.}
  The robustness of our scheme against rotation angle errors.
  The fidelity $F_{2}(t)$ for different rotation angle errors $\emph{d}\theta(t)$.
  The dashed-green, dotted-blue and solid-red lines correspond to $\eta=0.04\pi$, $0.06\pi$ and $0.08\pi$, respectively.
  Here $N=10$, $\tau=10^{-3}$, $B_{x}=4$, $B_{y}=5$, $B_{z}=6$.}
\end{figure}
\section*{APPENDIX C: Quantum Cram\'{e}r-Rao bound }
According to the parameter quantum estimation theory, the precision of the parameter is constrained by the quantum Cram\'{e}r-Rao bound (QCRB)~\cite{Helstrom,Helstrom1967,Paris2009},
\begin{eqnarray}\label{Eq5}
  \Delta B_{\alpha} \geq \Delta B_{\alpha}^\textrm{Q}\equiv \frac{1}{\sqrt{\eta F_{B_{\alpha}}^{Q}}}  \quad(\alpha=x,y,z),
\end{eqnarray}
which is characterized by the quantum Fisher information (QFI),
\begin{eqnarray}\label{Eq6}
  F_{B_{\alpha}}^{Q}=4\left(\langle \partial_{B_{\alpha}}\Psi |\partial_{B_{\alpha}}\Psi\rangle-|\langle \partial_{B_{\alpha}}\Psi|\Psi\rangle|^2\right).\nonumber\\
\end{eqnarray}
Here $\eta$ corresponds to the number of trials, $|\partial_{B_{\alpha}}\Psi\rangle$ denotes the partial derivative of $|\Psi\rangle$ with respect to the parameter ${B_{\alpha}}$.

In fact, particularly in experiments, one needs to find a suitable observable to approach the theoretical precision bounds.
According to the quantum estimation theory, the measurement precisions of the estimated parameters can be given according to the error propagation formula.
In our scheme, the measurement precisions for the three components of $B_{\alpha}$ via the population measurement are
\begin{equation}\label{Eq:Parameter uncertainty}
\Delta B_{\alpha}=\frac{(\Delta{\hat{J}_{z}})_{\text{f}}^{B_{\alpha}}}{|\partial{\langle\hat{J}_{z}\rangle_{\text{f}}^{B_{\alpha}}}/ \partial{B_{\alpha}}|},\quad (\alpha=x,y,z.)
\end{equation}
Here, $(\Delta{\hat{J}_{z}})_{\text{f}}$ and $\langle\hat{J}_{z}\rangle_{\text{f}}$ are respectively the standard deviation and expectation of $\hat{J}_z$ in the form of
\begin{equation}\label{Eq:Deviation}
(\Delta{\hat{J}_{z}})_{\text{f}}^{B_{\alpha}}=\sqrt{\langle\hat{J}_z^2\rangle_{\text{f}}^{B_{\alpha}}-(\langle{\hat{J}_{z}\rangle_{\text{f}}^{B_{\alpha}}})^2},
\end{equation}
with
\begin{equation}\label{Eq:Expectation}
\langle\hat{J}_{z}\rangle_{\text{f}}^{B_{\alpha}}=\!\bra{\Psi_{\text{final}}^{B_{\alpha}}} \hat{J}_{z} \ket{\Psi_{\text{final}}^{B_{\alpha}}},
\end{equation}
and
\begin{equation}\label{Eq:Expectation2}
\langle\hat{J}_{z}^2\rangle_{\text{f}}^{B_{\alpha}}=\!\bra{\Psi_{\text{final}}^{B_{\alpha}}} \hat{J}_{z}^2 \ket{\Psi_{\text{final}}^{B_{\alpha}}}.
\end{equation}
\section*{APPENDIX D:Derivation of the parallel scheme with Individual particles}
%
Here, we give the related derivation of the parallel scheme with individual particles.
In the first quantum interferometer, 
the final state before the half-population difference measurement is $|\Psi_{\text{final}}^{B_{x}}\rangle=e^{i\frac{\pi}{2}J_{x}}e^{-i\hat{H}_{B_{x}}^{\text{eff}}T_{x}}\ket{N/2,N/2}$.
In an explicit form, the final state becomes
\begin{eqnarray}\label{Evo_CSC1}
|\Psi_{\text{final}}^{B_{x}}\rangle=\frac{1}{2^{J}}\sum_{m=-J}^{m=J} {C_J^{m}}[\cos(B_{x}T_{x}/2)+\sin(B_{x}T_{x}/2)]^{J+m}\nonumber \\ \times[i\cos(B_{x}T_{x}/2)-i\sin(B_{x}T_{x}/2)]^{J-m}\ket{J,m},\nonumber \\
\end{eqnarray}
where $C_{J}^{m}={\frac{(2J)!}{(J+m)!(J-m)!}}$ is the binomial coefficient.
In the second quantum interferometer, 
the final state before the half-population difference measurement is $|\Psi_{\text{final}}^{B_{y}}\rangle=e^{i\frac{\pi}{2}J_{y}}e^{-i\hat{H}_{B_{y}}^{\text{eff}}T_{y}}\ket{N/2,N/2}$.
In an explicit form, the final state becomes
\begin{eqnarray}\label{Evo_CSC2}
|\Psi_{\text{final}}^{B_{y}}\rangle&&=\frac{1}{2^{J}}\sum_{m=-J}^{m=J} {C_J^{m}}[\cos(B_{y}T_{y}/2)+\sin(B_{y}T_{y}/2)]^{J+m}\nonumber \\ &&\times[i\cos(B_{y}T_{y}/2)-i\sin(B_{y}T_{y}/2)]^{J-m}\ket{J,m},\nonumber \\
\end{eqnarray}
In the last quantum interferometer, 
the final state before the half-population difference measurement is $|\Psi_{\text{final}}^{B_{z}}\rangle=e^{-i\frac{\pi}{2}J_{x}}e^{-i\hat{H}_{B_{z}}^{\text{eff}}T_{z}}e^{-i\frac{\pi}{2}J_{y}}\ket{N/2,N/2}$.
In an explicit form, the final state becomes
\begin{eqnarray}\label{Evo_CSC3}
|\Psi_{\text{final}}^{B_{z}}\rangle&&=\sum_{m=-J}^{m=J}{C_J^{m}}[e^{iB_{z}T_{z}/2}-ie^{-iB_{z}T_{z}/2}]^{J-m}\nonumber \\ &&\times[e^{-iB_{z}T_{z}/2}-ie^{iB_{z}T_{z}/2}]^{J+m}\ket{J,m}.
\end{eqnarray}
According to Eq.~(\ref{Eq6}), the QFI for the three parameter with the individual particles are $F_{B_{x}}^{\textrm{Q}}=NT_{x}^2$, $F_{B_{y}}^{\textrm{Q}}=NT_{y}^2$ and $F_{B_{z}}^{\textrm{Q}}=NT_{z}^2$.
Obviously, for all parameters $B_{\alpha}$, the ultimate precision bounds with individual particles under the parallel scheme can just attain the SQL (i.e. $\Delta B_{\alpha}^\textrm{Q} \propto 1/\sqrt{N}$).
In further, we consider the measurement precision via half-population difference measurement.
After some algebra, the expectations of half-population difference for the three quantum interferometries can be explicitly written as
\begin{eqnarray}\label{Jz_SCS1}
\langle J_{z} \rangle_{\text{f}}^{B_{\alpha}}=\frac{N}{2}\sin(B_{\alpha}T_{\alpha}),  \quad \alpha=x,y,z.
\end{eqnarray}
%
%
The square of half-population difference on the final state are
\begin{eqnarray}\label{Jz2_SCS1}
\langle J_{z}^2 \rangle_{\text{f}}^{B_{\alpha}}=\frac{N}{4}+\frac{N(N-1)}{4}\sin^2(B_{\alpha}T_{\alpha}),  \quad \alpha=x,y,z. \nonumber \\
\end{eqnarray}
%
%
According to Eq.~\eqref{Jz_SCS1}, it is found that the information of the estimated three parameters $B_\alpha$ can be inferred from the half-population difference.
Substituting Eq.~\eqref{Jz_SCS1} and Eq.~\eqref{Jz2_SCS1} into Eq.~\eqref{Eq:Parameter uncertainty}, one can analytically obtain $\Delta B_{\alpha}$.
They read as $\Delta B_{\alpha}={1}/{\sqrt{N} T_{\alpha}}$
\section*{APPENDIX E:Derivation of the parallel scheme with Entangled particles}
Here, we give the related derivation of the parallel scheme with entangled particles.
In the first quantum interferometer, 
the final state before the half-population difference measurement is $|\Psi_{\text{final}}^{B_{x}}\rangle=e^{-i\frac{\pi}{2}J_{z}^2}e^{i\frac{\pi}{2}J_{y}}e^{-i\hat{H}_{B_{x}}^{\text{eff}}T_{x}}e^{-i\frac{\pi}{2}J_{y}}|\textrm{GHZ}\rangle$.
In an explicit form, the final state becomes
\begin{small}
\begin{eqnarray}\label{Evo_GHZ1}
|\Psi_{\text{final}}^{B_{x}}\rangle&=&\frac{1}{2}[(1\!+\!i)\cos(\frac{NB_{x}T_{x}}{2})\!+\!(1+i)\sin(\frac{NB_{x}T_{x}}{2})]\!\ket{J,J}\nonumber\\
&+&\frac{1}{2}[(1\!+\!i)\cos(\frac{NB_{x}T_{x}}{2})\!-\!(1\!+\!i)\sin(\frac{NB_{x}T_{x}}{2})]\!\ket{J,-J}.\nonumber\\
\end{eqnarray}
\end{small}
%
In the second quantum interferometer, 
the final state before the half-population difference measurement is $|\Psi_{\text{final}}^{B_{y}}\rangle=e^{i\frac{\pi}{2}J_{z}^2}e^{-i\frac{\pi}{2}J_{x}}e^{-i\hat{H}_{B_{y}}^{\text{eff}}T_{y}} e^{-i\frac{\pi}{2}J_{x}}|\textrm{GHZ}\rangle$.
In an explicit form, the final state becomes
\begin{small}
\begin{eqnarray}\label{Evo_GHZ2}
|\Psi_{\text{final}}^{B_{y}}\rangle\!&=&\frac{1}{2}[(1\!+\!i)\cos(\frac{NB_{y}T_{y}}{2})\!+\!(1\!+\!i)\sin(\frac{NB_{y}T_{y}}{2})]\!\ket{J,J}\nonumber\\
&+&\frac{1}{2}[(1\!+\!i)\cos(\frac{NB_{y}T_{y}}{2})\!-\!(1\!+\!i)\sin(\frac{NB_{y}T_{y}}{2})]\!\ket{J,-J}.\nonumber\\
\end{eqnarray}
\end{small}
In the last quantum interferometer, 
the final state before the half-population difference measurement is $|\Psi_{\text{final}}^{B_{z}}\rangle=e^{-i\frac{\pi}{2}J_{x}}e^{i\frac{\pi}{2}J_{z}^2}e^{i\frac{\pi}{2}J_{x}}e^{-i\hat{H}_{B_{z}}^{\text{eff}}T_{z}}|\textrm{GHZ}\rangle$.
In an explicit form, the final state becomes
\begin{eqnarray}\label{Evo_GHZ3}
|\Psi_{\text{final}}^{B_{z}}\rangle&=&\frac{1}{2}[\cos(\frac{NB_{z}T_{z}}{2})+\sin(\frac{NB_{z}T_{z}}{2})]\ket{J,J}\nonumber \\
&+&\frac{1}{2}[\cos(\frac{NB_{z}T_{z}}{2})-\sin(\frac{NB_{z}T_{z}}{2})]\ket{J,-J}.\nonumber\\
\end{eqnarray}
According to Eq.~(\ref{Eq6}), the QFI for the three parameter in the parallel scheme with the entangled particles are $F_{B_{\alpha}}^{\textrm{Q}}=N^2T_{\alpha}^2$.
%
Obviously, for all parameters $B_{\alpha}$, the ultimate precision bounds with individual particles can just attain the HL (i.e. $\Delta B_{\alpha}^\textrm{Q} \propto 1/{N}$).
In further, we consider the measurement precision via half-population difference measurement.
After some algebra, the expectations of half-population difference for the three quantum interferometries can explicitly written as
%

\begin{eqnarray}\label{Jz_GHZ1}
\langle J_{z} \rangle_{\text{f}}^{B_{\alpha}}=\frac{N}{2}\sin(NB_{\alpha}T_{\alpha}),  \quad \alpha=x,y,z.
\end{eqnarray}
%
%
Clearly, the main frequencies of the bisinusoidal oscillation of half-population difference both become proportional to $N = 2J$.
The square of half-population difference on the final state  both are
\begin{eqnarray}\label{Jz2_GHZ}
\langle J_{z}^2 \rangle_{\text{f}}^{B_{\alpha}}=\frac{N^2}{4}.
\end{eqnarray}
Substituting Eq.~\eqref{Jz_GHZ1} and  Eq.~\eqref{Jz2_GHZ} into Eq.~\eqref{Eq:Parameter uncertainty}, one can analytically obtain $\Delta B_{\alpha}$ and they read as $\Delta B_{\alpha}={1}/{{N} T_{\alpha}}$.
%

\section*{APPENDIX F:Derivation of the sequential scheme with individual particles}
Here, we give the related derivation of the sequential scheme with individual particles.
Suppose all the particles are prepared in the spin coherent state (SCS) $\ket{\Psi}_{\textrm{SCS}}=\ket{N/2,N/2}$.
Then, the final state before the half-population difference measurement can be written as
\begin{eqnarray}\label{Evo_CSC}
|\Psi_{\text{final}}\rangle\!=e^{-i\frac{\pi}{2}{\hat{J}_{y}}} e^{-i\hat{H}_{B_{z}}^{\text{eff}}T_{z}} e^{-i\hat{H}_{B_{y}}^{\text{eff}}T_{y}} e^{-i\hat{H}_{B_{x}}^{\text{eff}}T_{x}}\ket{\textrm{SCS}}.\nonumber\\
\end{eqnarray}
In an explicit form, the final state becomes
\begin{widetext}
\begin{small}
\begin{eqnarray}\label{Evo_CSC}
|\Psi_{\text{final}}\rangle
&&=\!\!\!\!\sum_{m=-J}^{J}\!\!\!\!\frac{\sqrt{C_{J}^{m}}}{2^{J}}\!\left[\left(\!\!\cos(\!\frac{B_{x}\!T_{x}}{2}\!)\!\cos(\!\frac{B_{y}T_{y}}{2}\!)\!+\!i\sin(\!\frac{B_{x}\!T_{x}}{2}\!)\!\sin(\!\frac{B_{y}\!T_{y}}{2}\!)\!\right) \!e^{\!\frac{-i \!B_{z}\!T_{z}}{2}}\!\!-\!\!\left(\!\!\cos(\!\frac{B_{x}\!T_{x}}{2}\!)\!\sin(\!\frac{B_{y}\!T_{y}}{2}\!)\!\!-\!i\!\sin(\!\frac{B_{x}\!T_{x}}{2}\!)\!\cos(\!\frac{B_{y}\!T_{y}}{2})\!\!\right)\!e^{\!\frac{i \!B_{z}\!T_{z}}{2}\!}\right]^{J\!+\!m}\nonumber\\
&&\times\!\!\left[\!\left(\!\!\cos(\!\frac{B_{x}\!T_{x}}{2})\!\cos(\!\frac{B_{y}\!T_{y}}{2})\!+\!i\sin(\!\frac{B_{x}\!T_{x}}{2})\!\sin(\!\frac{B_{y}\!T_{y}}{2})\!\!\right)\! e^{ \frac{-i \!B_{z}\!T_{z}}{2}}\!+\!\left(\!\!\cos(\frac{B_{x}\!T_{x}}{2})\!\sin(\!\frac{B_{y}\!T_{y}}{2})\!-\!i\sin(\!\frac{B_{x}\!T_{x}}{2})\!\cos(\!\frac{B_{y}\!T_{y}}{2})\!\!\right)\!e^{ \frac{i\!B_{z}\!T_{z}}{2}}\!\right]^{J\!-\!m}\!\!|J,m\rangle, \nonumber\\
\end{eqnarray}
\end{small}
\end{widetext}
where $C_{J}^{m}={\frac{(2J)!}{(J+m)!(J-m)!}}$ is the binomial coefficient.
According to Eq.~(\ref{Eq6}), the QFI for the three parameter with SCS can be written as,
\begin{eqnarray}\label{QFI1_SCS}
F_{B_{x}}^{\textrm{Q}}=NT_{x}^2,
\end{eqnarray}
\begin{eqnarray}\label{QFI2_SCS}
F_{B_{y}}^{\textrm{Q}}= NT_{y}^2\cos^2(B_{x}T_{x}),
\end{eqnarray}
\begin{eqnarray}\label{QFI3_SCS}
F_{B_{z}}^{\textrm{Q}}= NT_{z}^2(1-\cos^2(B_{x}T_{x})\cos^2(B_{y}T_{y})).
\end{eqnarray}

Obviously, for all parameters $B_{\alpha}$, the ultimate precision bounds with individual particles can just attain the SQL (i.e. $\Delta B_{\alpha}^\textrm{Q} \propto 1/\sqrt{N}$).
In further, we consider the measurement precision via half-population difference measurement.
After some algebra, the expectations of half-population difference on the final state can be explicitly written as
\begin{small}
\begin{eqnarray}\label{Jz_SCS}
\langle J_{z} \rangle_{\text{f}}&=&\frac{N}{4}\![\cos(B_{x}T_{x}\!+\!B_{z}T_{z})\!-\!\cos(B_{x}T_{x}\!-\!B_{z}T_{z}\!)] \nonumber \\ &-&\frac{N}{8}\![\sin(B_{x}T_{x}\!+\!B_{y}T_{y}\!+\!B_{z}T_{z})\!+\!\sin(B_{x}T_{x}\!+B_{y}T_{y}\!-\!B_{z}T_{z}\!)]\nonumber \\
&+& \frac{N}{8}\![\sin(\!B_{x}T_{x}\!-\!B_{y}T_{y}\!+\!B_{z}T_{z}\!)\!+\!\sin(\!B_{x}T_{x}\!-\!B_{y}T_{y}\!-B_{z}T_{z}\!)].\nonumber\\
\end{eqnarray}
\end{small}
And the square of half-population difference on the final state can be explicitly written as
\begin{widetext}
\begin{eqnarray}\label{Jz2_SCS}
\langle J_{z}^2 \rangle_{\text{f}}&&=\frac{N}{4}+\frac{N(N-1)}{4}[\cos(B_{x}T_{x})\sin(B_{y}T_{y})\cos(B_{z}T_{z})-\sin(B_{x}T_{x})\sin(B_{z}T_{z})]^2.
\end{eqnarray}

Substituting Eq.~\eqref{Jz_SCS} and Eq.~\eqref{Jz2_SCS} into Eq.~\eqref{Eq:Parameter uncertainty}, one can analytically obtain $\Delta B_{\alpha}$.
They are read as,
\begin{eqnarray}\label{DeltaB1SCS}
\Delta B_{x}=\frac{1}{\sqrt{N} T_{x}}\frac{\sqrt{1-G}}{\left| \sin(B_{x} T_{x}) \sin(B_{y} T_{y}) \cos(B_{z} T_{z}) - \cos(B_{x} T_{x}) \sin(B_{z} T_{z}) \right|},\nonumber\\
\end{eqnarray}
\begin{eqnarray}\label{DeltaB2SCS}
\Delta B_{y}=\frac{1}{\sqrt{N}T_{y}}\frac{\sqrt{1-G}}{\left|\sin(B_{x}T_{x})\cos(B_{y}T_{y})\cos(B_{z}T_{z}))\right|},
\end{eqnarray}
and
\begin{eqnarray}\label{DeltaB3SCS}
\Delta B_{z}=\frac{1}{\sqrt{N}T_{z}}\frac{\sqrt{1-G}}{\left|\cos(B_{x}T_{x})\sin(B_{y}T_{y})\cos(B_{z}T_{z})-\sin(B_{x}T_{x})\sin(B_{z}T_{z})\right|},\nonumber\\
\end{eqnarray}
with
\begin{eqnarray}\label{G}
G=[\cos(B_{x}T_{x})\sin(B_{y}T_{y})\cos(B_{z}T_{z})+\sin(B_{x}T_{x})\sin(B_{z}T_{z})]^2. \nonumber\\
\end{eqnarray}
According to Eq.~\eqref{DeltaB1SCS} $\sim$ Eq.~\eqref{DeltaB3SCS}, the measurement precisions $\Delta B_{\alpha}$ for individual particles only exhibit the SQL scaling.

\section{APPENDIX G: Derivation of the sequential scheme with Entangled particles}

Here, we give the related derivation of the sequential scheme with entangled particles.
According to the main text, after the sequence, the final state can be written as

\begin{eqnarray}\label{Evo_Spin_cat_state3}
\ket{\Psi_{\text{final}}}&&= e^{i\frac{\pi}{2}{\hat{J}_{x}^{2}}} e^{-i\hat{H}_{B_{z}}^{\text{eff}}T_{z}}e^{i\frac{\pi}{2}{\hat{J}_{x}^2}}e^{-i\frac{\pi}{2}{\hat{J}_{x}}} e^{-i\hat{H}_{B_{y}}^{\text{eff}}T_{y}}e^{i\frac{\pi}{2}{\hat{J}_{z}^{2}}}e^{-i\frac{\pi}{2}{\hat{J}_{z}}} e^{-i\hat{H}_{B_{x}}^{\text{eff}}T_{x}}
\ket{\textrm{GHZ}}.
\end{eqnarray}
When $N$ is an even number, the final state $\ket{\Psi_{\text{final}}}$ has an analytic form that is written as
\begin{eqnarray}\label{Evo_GHZeven}
|\Psi\rangle_{\text{final}}^{I}=A_{1}\ket{J,J}+A_{2}\ket{J,-J}.
\end{eqnarray}
Here, the two coefficients read
\begin{eqnarray}\label{C_A1}
A_{1}=&&(-1)^{J}(e^{-iJ(B_{x}T_{x}+B_{y}T_{y}+B_{z}T_{z})}-ie^{iJ(B_{x}T_{x}+B_{y}T_{y}+B_{z}T_{z})}
-e^{-iJ(B_{x}T_{x}+B_{y}T_{y}-B_{z}T_{z})}-ie^{iJ(B_{x}T_{x}+B_{y}T_{y}-B_{z}T_{z})})\nonumber \\
&&-e^{-iJ(B_{x}T_{x}-B_{y}T_{y}+B_{z}T_{z})}+ie^{iJ(B_{x}T_{x}-B_{y}T_{y}+B_{z}T_{z})}
-e^{-iJ(B_{x}T_{x}-B_{y}T_{y}-B_{z}T_{z})}-ie^{iJ(B_{x}T_{x}-B_{y}T_{y}-B_{z}T_{z})},
\end{eqnarray}
\begin{eqnarray}\label{C_A2}
A_{2}=&&-ie^{-iJ(B_{x}T_{x}+B_{y}T_{y}+B_{z}T_{z})}+e^{iJ(B_{x}T_{x}+B_{y}T_{y}+B_{z}T_{z})}
-ie^{-iJ(B_{x}T_{x}+B_{y}T_{y}-B_{z}T_{z})}-e^{iJ(B_{x}T_{x}+B_{y}T_{y}-B_{z}T_{z})}\nonumber \\
&&+(-1)^{J}(ie^{-iJ(B_{x}T_{x}-B_{y}T_{y}+B_{z}T_{z})}-e^{iJ(B_{x}T_{x}-B_{y}T_{y}+B_{z}T_{z})}
-ie^{-iJ(B_{x}T_{x}-B_{y}T_{y}-B_{z}T_{z})}-e^{iJ(B_{x}T_{x}-B_{y}T_{y}-B_{z}T_{z})}).\nonumber \\
\end{eqnarray}
%
%
According to Eq.~(\ref{Eq6}), the QFI for the three parameter with GHZ can be written as,
\begin{eqnarray}\label{QFI1_GHZ}
F_{B_{x}}^{\textrm{Q}}=N^2T_{x}^2,
\end{eqnarray}
\begin{eqnarray}\label{QFI2_GHZ}
F_{B_{y}}^{\textrm{Q}}= N^2T_{y}^2[1+\cos(2NB_{x}T_{x})]/2,
\end{eqnarray}
and
\begin{eqnarray}\label{QFI3_GHZ}
F_{B_{z}}^{\textrm{Q}}= N^2T_{z}^2[1-\cos(NB_{x}T_{x})^2\sin^2(NB_{y}T_{y})].
\end{eqnarray}
%
The expectations of half-population difference on the final state can be written explicitly as
\begin{eqnarray}\label{Jz_GHZS}
\langle J_{z} \rangle_{\text{f}}=&&\frac{N}{8}[\sin((NB_{x}T_{x}+NB_{y}T_{y}+NB_{z}T_{z}))+\sin(NB_{x}T_{x}-NB_{y}T_{y}+NB_{z}T_{z})]\nonumber \\
&&-\frac{N}{8}[\sin(NB_{x}T_{x}+NB_{y}T_{y}-NB_{z}T_{z})+\sin(NB_{x}T_{x}-NB_{y}T_{y}-NB_{z}T_{z})]\nonumber \\
&&-(-1)^{J}\frac{N}{4}[\sin(N B_{x}T_{x}+NB_{z}T_{z})+\sin(NB_{x}T_{x}-NB_{z}T_{z})].
\end{eqnarray}
Moreover, the square of half-population difference is independent on the three parameters, i.e., s $\langle J_{z}^{2} \rangle_{\text{f}}={J^{2}}=N^2/4$.
According to Eq.~\eqref{Eq:Parameter uncertainty}, the measurement precisions for $B_{\alpha}$ can be analytically obtained:
\begin{eqnarray}\label{DeltaB1GHZS}
\Delta B_{x}=\frac{1}{{N}T_{x}}\frac{\sqrt{1-[\cos(NB_{x}T_{x})\cos(NB_{y}T_{y})\sin(NB_{z}T_{z})-(-1)^{J}\sin(NB_{x}T_{x})\cos(NB_{z}T_{z})]^2}}
{\left|\sin(NB_{x}T_{x})\cos(NB_{y}T_{y})\sin(NB_{z}T_{z})+(-1)^{J}\cos(NB_{x}T_{x})\cos(NB_{z}T_{z})\right|},
\end{eqnarray}
\begin{eqnarray}\label{DeltaB2GHZS}
\Delta B_{y}=\frac{1}{{N}T_{y}}\frac{\sqrt{1-[\cos(NB_{x}T_{x})\cos(NB_{y}T_{y})\sin(NB_{z}T_{z})-(-1)^{J}\sin(NB_{x}T_{x})\cos(NB_{z}T_{z})]^2}}
{\left|\cos(NB_{x}T_{x})\sin(NB_{y}T_{y})\sin(NB_{z}T_{z})\right|},
\end{eqnarray}
\begin{eqnarray}\label{DeltaB3GHZS}
\Delta B_{z}=\frac{1}{{N}T_{z}}\frac{\sqrt{1-[\cos(NB_{x}T_{x})\cos(NB_{y}T_{y})\sin(NB_{z}T_{z})-(-1)^{J}\sin(NB_{x}T_{x})\cos(NB_{z}T_{z})]^2}}
{\left|\cos(NB_{x}T_{x})\cos(NB_{y}T_{y})\cos(NB_{z}T_{z})+(-1)^{J}\sin(NB_{x}T_{x})\sin(NB_{z}T_{z})\right|}.
\end{eqnarray}
From Eqs.~\eqref{DeltaB1GHZ}, \eqref{DeltaB2GHZ} and \eqref{DeltaB3GHZ}, the measurement precisions $\Delta B_{\alpha}$ can exhibit Heisenberg-limited scaling(i.e.,$\Delta B_{\alpha}\propto \frac{1}{N}$).
\end{widetext}

\end{document}